\pdfoutput=1
\documentclass{cpbtex}
\usepackage{CJK}
\usepackage{lipsum}
\usepackage{graphicx, subfig}
\usepackage{footmisc}
\usepackage{hyperref}

\makeatletter
\def\@seccntformat#1{\csname the#1\endcsname.\ }
\makeatother

\newcommand{\s}{\mathbf{s}}
\newcommand{\calL}{\mathcal{L}}

\newcommand{\SHO}{\textbf{SHO}}
\newcommand{\AVE}{\textbf{AVE}}
\newcommand{\nMF}{\textbf{nMF}}
\newcommand{\TAP}{\textbf{TAP}}
\newcommand{\BM}{\textbf{BM}}

\begin{document}
\begin{CJK*}{GB}{gbsn}

\title{Inverse Ising techniques to infer underlying mechanisms from data \thanks{The work of H.-L. Zeng was supported partially by the National Natural Science Foundation of China (Grant No.~11705097), partially by Natural Science Foundation of Jiangsu Province (Grant No.~BK20170895), Jiangsu Government Scholarship for Overseas Studies of 2018 and Scientific Research Foundation of Nanjing University of Posts and Telecommunications (NY217013). The work of EA was partially supported by Foundation for Polish Science through TEAM-NET project (contract no. POIR.04.04.00-00-17C1/18-00).}
}


\author{Hong-Li Zeng(ÔøºìÀö)$^{1,2}$ \thanks{Corresponding author. E-mail:~hlzeng@njupt.edu.cn}\ ,
Erik Aurell$^{3,4}$\thanks{Corresponding author. E-mail:~eaurell@kth.se} \\
$^{1}${\small School of Science, Nanjing University of Posts and Telecommunications,}\\
{\small New Energy Technology Engineering Laboratory of Jiangsu Province, Nanjing 210023, China}\\
$^{2}${\small Nordita, Royal Institute of Technology, and Stockholm University, SE-10691 Stockholm, Sweden} \\
$^{3}${\small KTH -- Royal Institute of Technology, AlbaNova University Center, SE-106 91 Stockholm, Sweden}  \\
$^{4}${\small Faculty of Physics, Astronomy and Applied Computer Science, Jagiellonian University, 30-348 Krak\'ow, Poland}
}

\date{\today}
\maketitle

\begin{abstract}
As a problem in data science the inverse Ising (or Potts) problem is to infer the
parameters of a Gibbs-Boltzmann distributions of an Ising (or Potts) model
from samples drawn from that distribution.
The algorithmic and computational interest stems from the fact that
this inference task cannot be done efficiently by the maximum likelihood criterion, since
the normalizing constant of the distribution (the partition function) can not be calculated
exactly and efficiently.
The practical interest on the other hand flows from several outstanding applications,
of which the most well known has been predicting spatial contacts in protein structures
from tables of homologous protein sequences.
Most applications to date have been to data that has been produced by a dynamical
process which, as far as it is known, cannot be expected to satisfy detailed balance.
There is therefore no a priori reason to expect the distribution to be of
the Gibbs-Boltzmann type, and no a priori reason to expect that inverse Ising (or Potts)
techniques should yield useful information.
In this review we discuss two types of problems where
progress nevertheless can be made. We find that depending on model parameters
there are phases where, in fact, the distribution is close to Gibbs-Boltzmann distribution,
a non-equilibrium nature of the under-lying dynamics notwithstanding.
We also discuss the relation between inferred Ising model parameters
and parameters of the underlying dynamics.
\end{abstract}



\section{Introduction}
\label{se:introduction}
The Gibbs-Boltzmann distribution of the Ising model on $L$\protect\footnotemark[1]{$^1$}\protect\footnotetext[1]{$^1$ For later reference we prefer to refer to the number of spins in the model with the letter $L$, for ``loci''. The more customary letter $N$ will later be reserved to the number of samples drawn from the distribution, following a convention using in statistics.}
spins is
\begin{equation}\label{Gibbs-distribution}
    P(\s) = \frac{\exp\left(-\beta\left(\sum_i\theta_is_i+\sum_{i<j}J_{ij}s_is_j\right)\right)}{Z},
\end{equation}
where $\beta$ is the inverse temperature, and $Z$ is the partition function, defined as:
\begin{equation}\label{Partition-function}
    Z = \sum_{\s} \exp\left(-\beta \left(\sum_i\theta_is_i+\sum_{i<j}J_{ij}s_is_j\right)\right).
\end{equation}
The parameters of the model are $L$ external fields $\{\theta_i\}_{i=1}^L$ and
$\frac{L(L-1)}{2}$ coupling constants or interactions $\{J_{ij}\}_{i<j}$.
The Gibbs-Boltzmann distribution of a Potts model is defined in a similar way,
except that each variable can take $q$ values ($q=2$ for the Ising model)
and the model parameters are vectors and matrices\protect\footnotemark[2]{$~^2$}\protect\footnotetext[2]{$^2$ By reparametrization invariance the
number of independent paramaters is respectively $q-1$ for the vector and $(q-1)^2$ for the matrix,
which for $q=2$ gives only one parameter of each type as in \protect\eqref{Gibbs-distribution}.}
($\{\theta_i^{(\alpha)}\}$ for $1\leq\alpha\leq q$
and $\{J_{ij}^{(\alpha,\alpha')}\}$ for $1\leq\alpha,\alpha'\leq q$).

From the viewpoint of physics \eqref{Gibbs-distribution}
is the equilibrium distribution at inverse temperature $\beta$ corresponding
to the Ising energy function (or Hamiltonian) \cite{MezardParisiVirasoro,FischerHertz,MezardMontanari09}.
The traditional Ising problem of statistical mechanics is to determine
properties of the distribution $P(\s)$ from the model parameters $\{\theta_i,J_{ij}\}$.
The probability distribution $P(\s)$, or ensemble,
will be reflected in samples drawn independently from that distribution.
Combining the two steps of estimating the ensemble and sampling from the distribution,
the \textit{direct Ising problem} can be defined as the problem
of estimating an empirical probability distribution over samples
from model parameters.
The \textit{inverse Ising problem} is then the opposite problem
of inferring model parameters from samples drawn from the distribution \cite{Schneidman2006,Roudi-2009b,Nguyen-2017a}.

To stress the inverse nature of the problem it
is useful to introduce some notation from statistics.
The class of distributions \eqref{Gibbs-distribution},
with values of the external fields and interactions in some set, is called an \textit{exponential family}~\protect\footnotemark[3]{$^3$}
\protect\footnotetext[3]{$^3$ Exponential because the
parameters all appear in the exponent, and family because a set of parameters
are considered.}.
The inverse Ising problem is accordingly called
\textit{parameter inference in an exponential family} \cite{Wainwright-2008a}.
The most basic way to infer parameters from independent samples
from one and the same probability distribution is
\textit{maximum likelihood} (ML). For computational reasons ML is
often formulated in logarithmic coordinates as \textit{maximum log-likelihood}.
Given $N$ independent samples from \eqref{Gibbs-distribution}
maximum log-likelihood amounts to the convex optimization problem
\begin{equation}\label{ML}
\{\theta_i^*,J_{ij}^*\}^{ML}
=\hbox{arg}\hbox{max}\left[-\sum_i\theta_i\left<s_i\right>-\sum_{i<j}J_{ij}\left<s_is_j\right>-\frac{1}{\beta}\log Z\right]
\end{equation}
where $\left<s_i\right>$ and $\left<s_is_j\right>$ are the empirical averages computed from the samples.
The star on the parameters on the left-hand side mark that these are
\textit{inferred}, and the superscript $ML$ indicates the inference method.
The only reason \eqref{ML} is a difficult task is that the forward problem of computing $Z$
from the parameters is difficult.
The effect of the parameter $\beta$ cannot be separated from an overall scale of
$\{\theta_i^*\}$ and $\{J_{ij}^*\}$, and therefore only appears in \eqref{ML}
as a proportionality of the log-partition function $\log Z\left(\beta,\{\theta_i^*\},\{J_{ij}^*\}\right)$.
From now on we will, when not specified otherwise, set $\beta$ equal to one.

A fundamental fact of statistical inference, which holds for all exponential families,
is that maximum likelihood does not need all the data. Indeed, in \eqref{ML} data
only appear as empirical averages. That is, if we have a table of $N$ independent
samples this means $NL$ data items, but \eqref{ML} only depends on
$\frac{L(L+1)}{2}$ numbers computed from the data. Those numbers (here means and correlations)
are called \textit{sufficient statistics} for inference in an
exponential family~\cite{darmois35,koopman36}.
A second fundamental fact is that maximum likelihood inference gives the same result
as maximizing Shannon entropy conditioned by the sufficient statistics.
From the physical point of view this follows directly from
\eqref{Gibbs-distribution} being an equilibrium distribution,
which minimizes free energy.
Maximizing Shannon entropy conditioned by some chosen set of empirical
averages is called the \textit{maximum-entropy}~\cite{Jaynes1957,Aurell2016,Nimwegen-2016a}
or \textit{max-entropy}
approach to statistical inference. By the above such a set of
empirical averages is in one-to-one relation with a set of parameters
in an exponential family for which they are sufficient statistics.
This relation between exponential parameters and empirical averages
is called \textit{conjugacy}, or, in Information Geometry~\cite{Amari87,Amari2000}, a \textit{duality}.
The max-entropy approach with a given set of empirical averages is
equivalent to maximum likelihood inference in an exponential family
with the conjugate parameters.

In Physics \eqref{Gibbs-distribution} appears as a (canonical) equilibrium
distribution of a system interacting with a heat bath. Let two
configurations of the system be $\s$ and $\s'$, and let the probability of the system
to make the change
from $\s$ to $\s'$ per unit time be $W_{\s,\s'}$. Then equilibrium is reached
if the transition rates satisfy the \textit{detailed balance conditions}~\cite{VanKampen}
\begin{equation}\label{detailed-balance}
P(\s) W_{\s,\s'} = P(\s') W_{\s',\s}
\end{equation}
In equilibrium transitions from $\s$ to $\s'$ and $\s'$ to $\s$
are equally likely. As a consequence there
cannot be chains of states such that cyclic transitions in
one direction ($\s^{1}\rightarrow \s^{2}\rightarrow \cdots \rightarrow \s^{k} \rightarrow \s^{1}$)
is more likely than in the opposite direction
($\s^{1}\rightarrow \s^{k}\rightarrow \cdots \rightarrow \s^{2} \rightarrow \s^{1}$).
Chemistry and Biology have many examples of such cycles appear,
from chemical oscillations of the Belouzov-Zhabotinsky type to
the cell cycle and circadian rythms~\cite{Kuramoto,Goldbeter}. This immediately says that not
all dynamics on discrete state spaces can satisfy detailed balance,
and so cannot be expected to have stationary distributions like \eqref{Gibbs-distribution}.

If we focus on \textit{single-spin flips} and $P(\s)$ in \eqref{Gibbs-distribution}
we can write the detailed balance conditions as a relation between
\textit{spin flip rates} $r_i(+,\s_{\setminus i})$
and $r_i(-,\s_{\setminus i})$
\begin{equation}\label{detailed-balance-spin-flip}
r_i(-,\s_{\setminus i}) = r_i(+,\s_{\setminus i})e^{-2\beta \theta_i -2\beta \sum_j J_{ij} s_j}
\end{equation}
where $r_i(-)$ is the rate of spin $i$ to flip from down to up,
and $r_i(+)$ is the rate up to down. Both of them depend
on the configurations of all the other spins, written $\s_{\setminus i}$.
Alternatively we can write \eqref{detailed-balance-spin-flip}
as
\begin{equation}\label{detailed-balance-spin-flip-2}
r_i(\s) = \gamma_i(\s_{\setminus i}) e^{-\beta \Delta_i E(\s)}
\end{equation}
where $r_i(\s)$ is the rate of flipping spin $i$ in configuration $\s$,
$\Delta_i E(\s)$ is the energy change when doing so,
and  $\gamma_i(\s_{\setminus i})$ is an overall rate which
does not depend on the value of spin $i$.
Different \textit{Monte Carlo procedures} (or
Markov chain Monte Carlo (MCMC) algorithms) differ
by this overall rate $\gamma_i(\s_{\setminus i})$.

To give an example of a spin-flip dynamics which
does not satisfy detailed balance we point to the class of
\textit{focused algorithms} for \textit{constraint satisfaction problems},
invented by Christos Papadimitriou now three decades ago~\cite{Papadimitriou91,walksat,Barthel03,AurellGordonKirkpatrick,FMS,Alava08,KaSe07,Lemoy15,CME}.
In such algorithms one imagines that the energy function is
a sum of local terms all of which are one or zero. A solution
is a configuration where all the energy terms are zero (zero-energy ground state).
A focused algorithm is one where the rate of flipping spin $i$ is zero unless
at least one of the constraints depending on $i$ is unsatisfied, but otherwise
the dynamics remains partly random.
It is clear that for such dynamics one can flip into a satisfied
state, but once there the dynamics stops; one cannot flip out \protect\footnotemark[4]{$^4~$}
\protect\footnotetext[4]{$^4~$ The first condition
of focusing can be satisfied in the equilibrium algorithm \protect\eqref{detailed-balance-spin-flip-2}
by taking $\beta$ to infinity (zero temperature). But then the algorithm
is a deterministic greedy search, and is no longer random.}.
It is well known that focused algorithms such as ``walksat'' outperform equilibrium
algorithms in many important applications~\cite{walksat,KaSe07}.

Let us now go back to the problem of inferring the parameters
of the Ising model in \eqref{Gibbs-distribution} where
the data has been generated by some process which may or may
not satisfy detailed balance. The inference procedure is at
this point treated as a black-box. What does this mean?
Does it even make sense? When does it make sense?

In equilibrium statistical mechanics the answer is clear and simple:
the process makes sense if the data was generated by a process
in detailed balance with an energy function in the same
exponential family, and in a phase where sampling is possible.
The first condition simply means that if the data was generated
from a process with, say, third-order interactions between the spins,
those interactions will not be recovered
from inferring
only first-order and second-order interactions.
The second conditions means that parameters have to be such that the
dynamics explores enough configurations that there is enough
information to infer from.
A trivial example when this is not the case is zero temperature
where the configuration goes to a local minimum of the energy, and then does
not change.
A more subtle example is a spin glass phase where for large but not
infinite $\beta$ only part of the Gibbs distribution \eqref{Gibbs-distribution} will
be sampled by an MCMC algorithm unless the simulation time is exponentially large in system size~\cite{MezardParisiVirasoro}.
Inference from naturally generated samples, that are ``stuck in one valley'',
have long been known to be impossible by the class of inverse Ising methods surveyed here~\cite{Frontiers}.
For specific problems and with more tailored methods such a task
is sometimes nevertheless possible~\cite{Braunstein2011}.
Inference from samples
that are drawn uniformly from such a distribution has on the other hand been shown to be possible, and
even easy~\cite{ChauBerg2012}.
Such uniform samples however have to be generated by methods that either needs a large computational effort (long simulation time), or one needs to restart the simulation many times with new random initial values, which corresponds
to real data from many separate time series.

Once we step out of the realm of equilibrium dynamics we are much more in the dark.
For the specific example of Symmetric Simple Exclusion Process (SSEP)
it is known that the stationary distribution, \textit{i.e.}
the equivalent of \eqref{Gibbs-distribution}, contains all interactions
of all orders~\cite{Bertini2002,Derrida_2007},
meaning all single-spin and pair-wise terms as in \eqref{Gibbs-distribution},
all three-spin interactions, and so on.
 This is so even though
the SSEP dynamics is entirely specified by nearest-neighbor pairwise exclusion, and the non-equilibrium
aspects are only the boundary conditions, particle exchanges with reservoirs.
When the dynamics can be described
as depending on energy changes with some non-equilibrium element such
as focusing at every step (``bulk driven non-equilibrium process''),
the possibilities for the stationary distributions are wider still.
The outcome of an inverse Ising procedure applied to such data
may therefore be completely unrelated to the parameters of the mechanisms that
gave rise to the data.
The computational complexity
and number of data required to infer the parameters of any kind of non-equilibrium
steady state from snapshots has been shown to be daunting~\cite{DettmerChauBerg2016,Berg-2017a,DettmerBerg2018}.
Nevertheless, this is the setting of most successful
and interesting applications of inverse Ising techniques to date~\cite{Tkacic2014,Cocco-2018a}.
Why is this?

In this review we will present two cases where the above problem can
be analyzed and/or studied in simulations.
The first case is kinetic Ising models with possibly
different values of pairwise parameters $J_{ij}$ and $J_{ji}$.
When $J_{ij}=J_{ji}$ (symmetric kinetic Ising models) this is nothing
by a Monte Carlo procedure to compute the distribution $P(\s)$ in
\eqref{Gibbs-distribution}.
Models where $J_{ij}\neq J_{ji}$ (asymmetric kinetic Ising models)
have however also been widely studied, \textit{e.g.}
as model systems in neuroscience~\cite{Parisi86,Frontiers,RoudiTyrchaHertz2009}.
The kinetic Ising models hence interpolate between equilibrium and
non-equilibrium systems. They also illustrate that more efficient
inference procedures than inverse Ising are available if one
can use a time series and not only independent samples from
a stationary distribution.

The second case are slightly more involved spin dynamics that model
evolution under mutations, Darwinian selection (fitness),
finite-$N$ effects (genetic drift) and recombination (sex).
We will here see that inverse Ising works in certain ranges
of parameters describing the relative strengths of
mutations, fitness and sex, but not in others.
We will also see that the relation is not trivial;
non-trivial theory is needed to translate the results from inverse Ising
to inferred fitness that can be compared to model parameters.

This review is organized as follows.
In Section~\ref{techniques},
we summarize for completeness
some inverse Ising techniques.
This topic is already covered by excellent reviews
to which we refer for more details and a wider palette
of techniques.
In Section~\ref{sec:kinetic-Ising} we introduce the
kinetic Ising problem in its symmetric and asymmetric
form, and present characteristic results,
and in Section~\ref{sec:applications-Ising}, we present
two applications of those techniques taken from
earlier work by one of us (HLZ).
Section~\ref{sec:population-genetics} presents on the other hand
a class of  problems in population genetics, and
Section~\ref{sec:discussion} contains an outlook
and discussion.

\section{Techniques for Inverse Ising}
\label{techniques}
The inverse Ising problem has been studied under several different names,
such as statistical inference in exponential families (as above), Boltzmann machines,
maximum-entropy modeling,
Direct Coupling Analysis (DCA), logistic regression techniques, and more.
For small enough system (small enough $L$)
maximum likelihood \eqref{ML} is computationally feasible,
for instance
by the iterative method also known as Boltzmann machine~\cite{Ackley1985}.
The idea of that very widely used method is to adjust the parameters in the
exponential family to make empirical averages and ensemble averages
of the conjugate sufficient statistics agree.

For large $L$ maximum likelihood (ML) is not computationally efficient,
meaning that it requires an effort exponentially increasing in $L$.
It should be said that for a given fixed $L$, what is and is not
computationally feasible changes with time and the development
of computer hardware. Nevertheless, for many applications that have been of interest, either ML
has not been feasible, or other inference schemes have given comparable
results with less effort.
In any case, it has been an interesting theoretical challenge to design
and analyze schemes that make a different trade-off between accuracy
and computational speed than ML.

The state of the art of inverse Ising was recently extensively reviewed in \cite{Nguyen-2017a},
and we will here only provide a background for the later sections.
A first type of inference methods attempts to circumvent
the computational challenge of ML by estimating
the partition function $Z$ efficiently.
Such methods are collectively known as \textit{mean-field inference},
because they rely on mean-field techniques.
The by far most common version of mean-field inference relies
on a variational ansatz in terms of magnetizations, which yields the
\textit{physical mean-field equations} of the Ising model
\begin{equation}
\label{eq:mean-field}
m_i = \tanh\left(h_i+\sum_j J_{ij} m_j\right)
\end{equation}
In this equation only $m_i$ is taken from the data, and there are only $L$
equations. By using also linear-response
\begin{equation}
\label{eq:linear-response}
c_{ij} = \left<s_is_j\right> - \left<s_i\right> \left<s_j\right> = \frac{\partial m_i}{\partial h_j}
\end{equation}
one finds the \textit{naive mean-field} inference formula~\cite{Rodriguez97efficientlearning}
\begin{equation}\label{eq:nMF_equi}
J_{ij}^{*,nMF} = - \left(c^{-1}\right)_{ij}
\end{equation}
The above expression is computationally quite convenient
as it reduces a complicated inference to matrix inversion.
One may note that \eqref{eq:nMF_equi} is the same formula as inferring
the interaction matrix of a Gaussian model
(precision matrix in information theory) from data.
It is an elementary property of multidimensional centered
Gaussian distributions that they can be
written $P(\mathbf{x})=\frac{1}{N}\exp\left(-\frac{1}{2}\mathbf{x}C^{-1}\mathbf{x}\right)$ where
$C$ is the co-variance matrix. The precision matrix (the model parameters)
can therefore be inferred as the inverse matrix of $C$ (the data).
The difference is that for an Ising model \eqref{eq:nMF_equi} is
only approximate, and does not always with good accuracy;
for the SK model (to be discussed below) it holds for instance
at high-temperature (weak interactions), but not
at low temperature.
If needed one can combine
\eqref{eq:mean-field} and \eqref{eq:nMF_equi}
to estimate also the external fields, \textit{i.e.}
\begin{equation}
\label{eq:nMF_h}
h_{i}^{*,nMF} = \tanh^{-1} m_i - \sum_j J_{ij}^{*,nMF} m_j
\end{equation}

More advanced mean-field methods than naive mean-field are obtained by starting from
more advanced approximations
than \eqref{eq:mean-field}. The best-known of these is TAP (Thouless-Anderson-Palmer)~\cite{TAP}
which starts from
\begin{equation}
\label{eq:TAP}
m_i = \tanh\left(h_i+\sum_j J_{ij} m_j - m_i \sum_j J_{ij}^2 (1-m_j^2) \right)
\end{equation}
Using linear response then gives $J_{ij}^{*}$ as the solution of a quadratic equation
\begin{equation}
\label{eq:TAP-formula}
J_{ij}^{*,TAP} + 2 m_i m_j \left(J_{ij}^{*,TAP}\right)^2 = - \left(c^{-1}\right)_{ij}
\end{equation}

A general feature of inference methods of this type is that in the variational
ansatz the data is only taken into account through the single-variables marginals,
\textit{i.e.} through the magnetizations. It is only linear-response \eqref{eq:linear-response},
which is a exact property of the full Ising model, but not of the variational ansatz,
that two-variable marginal are brought back into play.

Another type of mean-field inference equation attempts
to find the Ising model which best fits the data. The variational
parameters are then magnetizations ($m_i$) and correlations ($c_{ij}$),
conjugate to model parameters $h_i$ and $J_{ij}$.
This approach was
first developed as an iterative procedure called ``susceptibility propagation''~\cite{MezardMora,Weigt-2009a}
and only later shown to also yield equations
like \eqref{eq:nMF_equi} and \eqref{eq:TAP-formula}
where ratios of hyperbolic functions appear on the left -hand side,
but the right-hand side is still just the inverse matrix of correlations~\cite{Ricci_Tersenghi2012}.
An alternative derivation of this elegant approach can be found in~\cite{Nguyen-2017a},
which also contains a survey of many more methods that have been introduced and tested in the literature.

A different type of inference gives up on the ambition to
approximate the partition function, and hence the full probability distribution $P(\s)$.
Instead one tries to infer the parameters from some other property which can be efficiently computed.
The most widely used such method is
\textit{maximum pseudo-likelihood}~\cite{Besag-1975a}
or \textit{pseudo-likelihood maximization} (PLM).
This starts from the \textit{conditional probability} of the Ising model
\begin{equation}\label{Ising-conditional}
    P(s_i|\s_{\setminus i}) = \frac{\exp\left(-\beta\left(\theta_is_i+\sum_{j}J_{ij}s_is_j\right)\right)}
                           {\sum_{s'=\pm}\exp\left(-\beta\left(\theta_is'+\sum_{j}J_{ij}s's_j\right)\right)}
\end{equation}
In contrast to \eqref{Gibbs-distribution} there is now no longer any difficult
to compute normalization factor. The denominator of \eqref{Ising-conditional} is
the normalization of a distribution over only one Ising spin, and hence has only two terms.
When treated in the same way as ML \eqref{ML}, \eqref{Ising-conditional} leads to
$L$ inference problems, one for each spin $i$
\begin{equation}\label{PML}
\left(\theta_i^*,J_{ij}^*\right)^{PML,i}
=\hbox{arg}\hbox{max}\left[-\theta_i\left<s_i\right>-\sum_{j}J_{ij}\left<s_is_j\right>-\frac{1}{\beta}\left<\log \zeta_i\right>\right]
\end{equation}
where $\zeta_i$ is the sum in the denominator of \eqref{Ising-conditional}.
The left hand side emphasizes that this is inference ``as seen from spin $i$'' (by maximizing conditional
probability of spin $i$). To get the final answer one needs to combine
$J_{ij}^{*,PML,i}$  and $J_{ij}^{*,PML,j}$, typically by taking their average.

In the limit of infinite data PLM will almost surely find the same parameters as ML,
a property referred to as \textit{statistical consistency}
\protect\footnotemark[5]{$^5$}
\protect\footnotetext{$^5$ The formal definition of statistical consistency is
that as the number of samples goes to infinity, the argmin of the estimator converges in probability to the right answer. This holds for ML and PLM and some other inference methods to be discussed
below, but does not hold for mean-field inference methods.
In the limit of infinite data the sample averages used in mean-field
will always surely be the same as ensemble averages, but the recovered
parameters will not be the true ones because physical mean-field is
in itself approximate.
For a discussion, see \textit{e.g.}~\protect\cite{Nguyen-2017a} and references cited therein.
}.
In applications PLM has often been found to outperform both
naive and advanced mean-field inference~\cite{Nguyen-2017a}.
Why that is so cannot be said to be completely known,
since the number of samples in real data sets is finite.
The error of mean-field inference compared to
PLM in the infinite sample limit (lack of statistical consistency)
could therefore be compensated by the error in PLM
when used on a finite number of samples.
Empirically this has mostly not been found to be the case, but that
may partially be a consequence of the kinds of data sets
that have been considered in the literature.

\subsection{Undersampling, regularization, prior information and evaluation criteria}
\label{sec:undersampling}
\textit{High-dimensional statistics} is the branch of modern statistics
where the number of samples ($N$) is assumed to grow together with
or slower than the
number of parameters (here $\frac{L(L+1)}{2}$).
Common sense says that if there are fewer samples
than parameters and no other information, then the parameters cannot be fully determined by the
data.
This rule-of-thumb has to be applied with care, because often
there is other information, used explicitly or implicitly;
we will refer to a few such cases below.

Nevertheless, the rule-of-thumb points to something
important, namely that in the important application of inverse Potts methods
to contact prediction in protein structures~\cite{Morcos-2011a,Marks2011},
the number of parameters\protect\footnotemark[6]{$^6$}
\protect\footnotetext[6]{$^6~$ For $20$ types of
amino acids in a protein of $100$ residues.}is typically about $20^2 \cdot 100^2$, which is four million,
while the number of samples is rarely more than a hundred thousand.
All inference methods outlined above are therefore in
this application used in regimes where they are under-sampled, and so need to be regularized.
For naive mean-field inference  a regularization by pseudo-counts (adding fictitious
uniformly distributed samples) was used in~\cite{Morcos-2011a,Marks2011},
while an $L_1$-regularization was used in~\cite{Jones-2012a},
and an $L_2$-regularization in~\cite{Andreatta-2014a}.
For PLM similarly $L_2$-regularization was used in~\cite{Kamisetty-2013a} and
\cite{Ekeberg-2013a,Ekeberg-2014a}.

An important aspect of all inference is what is the family from which one tries
to infer parameters.
This can be given a Bayesian interpretation as an \textit{a priori}
distribution of parameters; the more one knows in that direction, the better the inference can be.
Many regularizers can be seen as logarithms of Bayesian prior distributions
such that the analogy also works the other way: regularized inference
is equivalent to inference with a prior (exponential of the regularizer), and can therefore work better
because it uses more information.
For instance, if all parameters are supposed
to be either zero or bounded away from zero by some lower threshold value, and if the ones
that are non-zero are sparse, then the authors of~\cite{Ravikumar-2010a}
showed that $L_1$-regularized PLM can find the graph structure
using relatively few samples, given certain assumption that were later
shown to be restrictive~\cite{BentoMontanari2009}. Nevertheless, using
and analyzing thresholding also in the retained predictions, the authors
of~\cite{Lokhov-2018a} were able to show that $L_1$-regularized PLM can
indeed find the graph structure using order of $\log L$
samples~\protect\footnotemark[7]{$^7$}
\protect\footnotetext{$^7$ The same authors also showed that $L_2$-regularized PLM
with thresholding, as used in the plmDCA software of \cite{Ekeberg-2014a}
can recover parameters in  $L_2$ norm using order of $\log L$ samples.}.

A second and equally important aspect is the evaluation criteria.
The criterion in~\cite{Ravikumar-2010a} is \textit{graphical}: the objective is to
infer properties of the model (the non-zero interactions) which can be represented as
a graph. It is obvious that inference under this criterion will be difficult
without a gap in the distribution of interaction parameters away from zero.
Information theory imposes limits on the smallest couplings that can be retrieved from the finite
amount of data~\cite{SanthanamWainwright2012,Lokhov-2018a};
given finite data it
is simply not possible to distinguish a parameter which is strictly zero from
one which is only very small.
Another type of criterion is \textit{metrical}, most often the squared differences of
the actual and inferred parameter values \cite{Frontiers,Nguyen-2017a}.
Yet another is \textit{probabilistic} by determining some difference
between the two probability distributions as in \eqref{Gibbs-distribution},
one with the actual parameters and one with the inferred parameters.
Two examples of probabilistic criteria are Kullback-Leibler divergence and
variational distance. An advantage of probabilistic criteria is that they focus
on typical differences of samples, and not on parameter differences which may
(sometimes) not matter so much as to the samples observed. However, as this
requires sampling from the distributions, it is also a disadvantage.

Important results have been obtained as to how many samples are
required for successful inference when both the a priori distributions
and and the criteria are varied, first in~\cite{Ravikumar-2010a}
under strong assumptions,
and more recently in~\cite{Vuffray-2016a,Lokhov-2018a}.
These two latter papers also introduce a different objective function, Interaction Screening
Objective (ISO), that has dependence on the same local quantities as pseudo-likelihood,
and which provably outperforms PLM in terms of expected error for the given number of samples,
providing near sample-optimal guarantee.
ISO has also more recently been generalized to
learn Ising models in the presence of samples corrupted by independent noise~\cite{pmlr-v99-goel19a},
and to the case of Potts models and beyond
pairwise interactions~\cite{Vuffray2019}.

In practice and in many successful applications to real data,
criteria have been of the type ``correctly recovering $k$ largest interactions'',
colloquially known as ``top-$k$''. Performance under such criteria
is straight-forward to analyze empirically when there is a known
answer; one simply compiles two lists of $k$ largest parameters
and what interactions they refer to, and then compares the two lists.
For instance, one can check what fraction of
$k$ largest inferred interactions can also be found among the
$k$ largest actual interactions, which is known as $k$-True Positive Rate, or $TPR(k)$.
In the application of inverse Potts methods
to contact prediction in protein structures~\cite{Morcos-2011a,Marks2011},
$k$ has commonly been taken to be around $100$.
The inequality that number of retained parameters be
less than the number of samples has hence been
respected, with a large margin.
The theoretical analysis of performance under this type of criterion is however more
involved, as the distribution of the largest values of
a random background is an extreme deviations problem.
One approach is to leverage an $L_{\infty}$ norm guarantee~\cite{Lokhov-2018a,Vuffray2019},
for another using large deviation theory, see \cite{Xu-2017a, Xu2018}.

\subsection{Time series and \textit{alltime} inverse Ising techniques}\label{sec:time-series}

In the following Section~\ref{sec:kinetic-Ising} we will consider
inference from data generated by a kinetic Ising
model, and in Section~\ref{sec:applications-Ising} we will consider
applications of this technique to data in Neuroscience and from Finance.
The main message of these sections will be that \textit{if you have time series data,
it is usually better to do inference on the time-labeled data}.
As we will show, even when the dynamics is of the type
\eqref{detailed-balance-spin-flip-2}, respects detailed balance,
and has stationary distribution \eqref{Gibbs-distribution}, it can be faster and easier to infer $J_{ij}$ from the dynamical law than by inverse Ising techniques.

Nevertheless, even if the data was generated in a dynamic process, we do not always have time series data.
In Sections~\ref{sec:population-genetics-dynamics} and ~\ref{sec:fitness-inference-synthetic} we will consider models of evolution, intended as stylized descriptions of the kind of genetic / protein data on which inverse Ising (Potts) techniques have been applied successfully~\cite{Morcos-2011a,Marks2011,Stein2015,Cocco-2018a}.
The underlying dynamics is then of the type of $N_{tot}$ individuals (genomes / genetic signatures / proteins)
of size (genomic length) $L$ evolving for a time $T$, while the data is on $N$ individuals (genomes / genetic signatures / proteins) sampled at one time\protect\footnotemark[8]{$^8$}
\protect\footnotetext[8]{$^8$ Or at uneven times so that the time information is hard to use, or the time at which they were sampled is unknown, the cases may differ depending on the data set.}.

Averages at any given time will have errors which go down as
$\left(N_{tot}\right)^{-\frac{1}{2}}$, typically a very
small number for real data sets, but not necessarily very small in a simulation.
For the evaluation of how simulations match theory
it is therefore of interest to also consider as
input data to inverse Ising variants of
naive mean-field \eqref{eq:nMF_equi} and PLM \eqref{PML}
where the averages are computed both over samples
and over time. We refer to these variants as \textit{alltime}
versions of the respective algorithms.

\section{A Model: Kinetic Asynchronous Ising Dynamics}
\label{sec:kinetic-Ising}
A standard approach to sample the equilibrium Ising model
is Glauber dynamics \cite{glauber1963time,suzuki1968dynamics}.
On the level of probability distributions
it is formulated as master equations
\begin{equation}\label{master_equation}
  \frac{d}{dt}p(s_1,...,s_L;t) = \sum_i \omega_i(-s_i)p(s_1,...,-s_i,...,s_L;t) -  \sum_i \omega_i(s_i) p(\s; t)
\end{equation}
where $\omega_i(s_i)$ is the flipping rate, \textit{i.e.}, the probability for the state of $i$th spin
to changes from $s_i$ to $-s_i$ per unit time while the other spins are momentarily unchanged.
Equation (\ref{master_equation}) shows that the configuration $s_1,...,s_L$ is destroyed by a flip of any spin $s_i$ (a \textit{loss term}), but it can also be created by the flip from any configuration with the form $s_1,...-s_i,...,s_L$ (a \textit{gain term}).
The flipping rate of spin $i$ is
\begin{eqnarray}\label{flipping_rate}
\begin{aligned}
 \omega_i(\s)&=\dfrac{\gamma}{1+\exp\left[2s_i\left(\theta_i+\sum_jJ_{ij}s_j\right)\right]}\\
 &=\frac{\gamma}{2}\left[1-s_i\tanh\left(\theta_i+\sum_j
 J_{ij}s_j\right)\right]
 \end{aligned}
\end{eqnarray}
The parameter $\gamma$ is an overall rate which in Glauber dynamics is assumed to
be the same for all spins.
The left-hand side depends on the whole configuration $\s$ because the values of all
spins enter on the right-hand side. The inverse temperature $\beta$ is here set to be $1$; as noted above it can be absorbed in the parameters.

For small enough systems (small $L$) \eqref{master_equation} can be simulated by solving $2^L$ linear ordinary differential equations.
For larger $L$ \eqref{master_equation} can only be simulated by Monte Carlo procedure. This
means that one considers $N$ separate spin configurations $\s_1,\ldots,\s_N$, each of which is evolved
in time. The empirical probability distribution
\begin{equation}\label{field-t}
 P^e(\s,t)=\frac{1}{N}\sum_{s=1}^N \mathbf{1}_{\s,\s_s(t)}
\end{equation}
is then an approximation of $P(\s,t)$ in \eqref{master_equation}.
We note (trivially) that for large systems $P^e(\s,t)$ will typically be either
zero or $\frac{1}{N}$; the chance that among $N$ separate spin configurations $\s_1,\ldots,\s_N$
two are exactly equal will be very small.
$P^e(\s,t)$ hence approximates $P(\s,t)$ as to certain summary statistics such as
single-spin averages (magnetizations), but typically cannot approximate $P(\s,t)$
very well as to the values for individual configurations.

For simplicity of presentation we will here focus on the time-homogeneous case where all parameters are
time independent. Distributions will then eventually relax to a stationary state, and we
will assume that this process has taken place. Inference can then by done
by treating samples at different times as independent, \textit{i.e.} by the type of \textit{alltime} algorithms discussed in Section~\ref{sec:time-series}.
For the rest of this section $N$ (the number of different time series)
will hence be one. Indeed, as in the Monte Carlo procedure the
different samples do not interact, one can limit oneself to
just one time series, as long as one is interested in properties of the statistically
stationary state reached at large times.

The dynamics of a configuration $\s(t)$ is governed by the same rates as
in \eqref{master_equation}.
In the Monte Carlo simulation scheme
it is convenient to consider spin $i$ as responding to an effective field from
the external field $\theta_i$ and the interactions from all the other spins.
This effective field is time-dependent, because the configurations of the
other spins change in time, \textit{viz.}
\begin{equation}\label{field}
 H_i(t)=\sum_jJ_{ij}s_j(t)+\theta_i.
\end{equation}
and the instantaneous rates are then
\begin{equation}\label{flipping_rate-with-H}
 \omega_i(\s,t)=\frac{\gamma}{2}\left[1-s_i(t)\tanh\left( H_i(t)\right)\right]
\end{equation}

One approach to simulation is to introduce a small
time step increment $\delta t$ and to flip each spin at each time with probability $\omega_i(\s,t)$.
For this scheme to simulate \eqref{master_equation} one must take $\delta t$ so small
that the chance of any other spin to flip in the same short time interval is negligible.
This scheme can be said to rely on $L\cdot t/\delta t$ random variables,
one for the decision whether or not to flip each spin in each time interval.
Since on average less than one spin will flip in each time interval
the probabilities of these variables have to be very biased towards not flipping.

A computationally more efficient scheme
is to first consider the rate of the event of flipping any spin.
That is
\begin{equation}
\omega_{TOT}(\s,t) = \sum_i \omega_i(\s,t)
\end{equation}
As long as no spin flips this overall rate does not change. The waiting time
until any spin has flipped is therefore an exponentially distributed random
variable with rate $\omega_{TOT}$, and the chance that it was spin $i$ that
flipped is $\omega_i/\omega_{TOT}$. The dynamics can then be simulated in discrete
steps starting from a configuration $\s_0$ at $t_0$
such that flips take place at times $t_1,t_2,\ldots$.
Initially the rates are $\{\omega_i(\s_0)\}$ and $t_1-t_0$
is an exponentially distributed random variable with rate $\omega_{TOT}(\s_0)= \sum_i \omega_i(\s_0)$.
The first spin to flip will be the $j$'th spin with probability
$\omega_j(\s_0)/\omega_{TOT}(\s_0)$, and after the flip all rates are updated
to $\{\omega_i(\s_1)\}$, and the process is repeated.
This algorithm is called the \textit{Gillespie algorithm}~\cite{Gillespie1977},
and relies on $L\cdot t/\overline{\Delta t}$ random variables
where $\overline{\Delta t}$ is some characteristic
time interval between the flips. At the price of a slightly more
complicated structure it is thus faster than the first algorithm by a ratio $\overline{\Delta t}/\delta t$.
Furthermore this method is exact; $\overline{\Delta t}$ is a property
of the dynamics and not of the simulation scheme.

A third approach is to update at each step a spin $i$ picked uniformly
at random with probability $\gamma\delta t$.
After such an update, which may or may not change the spin value,
the new value will be
      \begin{displaymath}
        s_i(t+\delta t) = \left\{
        \begin{array}{ll}
        +1 & \textrm{~~~~with probability~~~ $1/\{1+\exp[-2\beta H_i(t)]\}$}\\
        -1 & \textrm{~~~~with probability~~~ $1/\{1+\exp[2\beta H_i(t)]\}$}
        \end{array} \right.
      \end{displaymath}
From this we can evaluate the rate of flipping of spin $i$ per unit time to be
      \begin{displaymath}
       \left\{
        \begin{array}{ll}
        \gamma/\{1+\exp[2\beta H_i(t)]\}  & \textrm{when $s_i(t)=1$} \\
        \gamma/\{1+\exp[-2\beta H_i(t)]\} & \textrm{when $s_i(t)=-1$}
        \end{array} \right.
      \end{displaymath}
which gives the same rate as in \eqref{flipping_rate}.
Since two random numbers are called for each spin at each time interval, this scheme can
be said to rely on $2L\cdot t/\delta t$ random variables.

\subsection{Symmetric and asymmetric Sherrington-Kirkpatrick(SK) models}
\label{sec:symmetric-asymmetric-SK}
As illustrative examples we will now look at symmetric and asymmetric SK
models \cite{Sherrington1975} which are defined as follows.
First we introduce $J_{ij}$ with no restriction on $i$ and $j$.
Such a matrix can be split into its symmetric and asymmetric parts.
We write
\begin{equation}\label{Jij}
    J_{ij} = J_{ij}^s + kJ_{ij}^{as},~~~~~~~~~~~~~k \geq 0,
\end{equation}
where $J_{ij}^s$ and $J_{ij}^{as}$ are symmetric and asymmetric interaction respectively:
\begin{eqnarray}\label{Jij_s_as}
\begin{aligned}
    J_{ij}^s &= J_{j i}^s,\\
    J_{ij}^{as} &= -J_{ji}^{as}
\end{aligned}
\end{eqnarray}
The parameter $k$ in equation (\ref{Jij}) measures the asymmetric degree of the interactions $J_{ij}$. With $k=0$, $J_{ij}$'s are a fully symmetric model
the stationary distribution of which is \eqref{Gibbs-distribution}.
Any $k\neq 0$ means the $J_{ij}$ and $J_{ji}$ are not the same, and we have a non-equilibrium dynamics.
The SK kinetic model, extended to non-equilibrium \cite{crisanti1987},
means to take both the symmetric and the asymmetric couplings to be identically and independently Gaussian distributed random variables with
means zero and variances
\begin{equation}\label{variables_J_s_as}
  \langle{J_{ij}^s}^2\rangle = \langle{J_{ij}^{as}}^2\rangle =\dfrac{g^2}{N}\frac{1}{1+k^2}.
\end{equation}
This parametrization is chosen such that the total coupling matrix $J$ follows a Gaussian distribution
\begin{equation}\label{variables_J}
 p\left(J_{ij}\right) \varpropto \exp\left(-\frac{\left(J_{ij}-\mu\right)^2}{2\sigma^2}\right)
\end{equation}
with means $\mu=0$ and variance $\sigma^2 = g^2/N$ independently of $k$.

The interactions $J_{ij}$ define spin update rates \eqref{flipping_rate}
or \eqref{flipping_rate-with-H}. To see that asymmetric interactions
do not lead to Gibbs distributions \eqref{Gibbs-distribution}, it is useful
to temporarily change the parametrization so that there are three
only non-zero interactions $J_{ij}=J_{jk}=J_{ki}=J$, all large.
All other $J_{ij}$ are zero, and
all $\theta_i$ are also zero.
Assume that initially the three spins $s_i$, $s_j$ and $s_k$ are all
up \textit{i.e.} $+++$. They will then have the same (small) flip
rate $\gamma/\left(1+e^J\right)$, and one of them will flip first, let that be spin $i$,
so that the next state is $-++$.
After this has happened the (much larger) rate for either $i$ to flip back or for $k$ to flip
will be $\gamma/\left(1+e^{-J}\right)$. A flip of spin $i$ will hence almost surely
either go back to the starting state $+++$ after two flips, or lead to the configuration $-+-$.
This second state will in turn almost surely lead to
$++-$ or $---$.
The first of these is a shift of the state after the first flip
to the left, and by circular permutation symmetry it must be more likely
that the shifts continue in that direction rather than to the right.
The second is on the other hand obviously the mirror image of the starting state,
and all rates are again low.
Flipping out of $---$ would lead to $+--$, which would give $+-+$
and then $+++$ or $--+$, which is also a shift to the left.
A dynamics which has some similarities
to the above where
motion surely goes only in one direction is the basis of
Edsger Dijkstra's famous self-stabilizing system under distributed
control~\cite{Dijkstra1974}, for a physics perspective, see~\cite{Aurell_2013}.

\subsection{Inference for asynchronous Ising models}\label{sec:asynchronous-Ising-models}
Many techniques for inverse Ising as discussed above in Section~\ref{techniques}
have been applied to data from asynchronous Ising (or similar) dynamics,
mainly for neuroscience applications~\cite{Schneidman2006,RoudiJoannaHertz2009,shlens2006Neuroscience,cocco2009neuronal}.
Since our purpose here is to compare to inference using a time series
we will for the equilibrium case just consider the simplest method, which
is naive mean-field (nMF) \eqref{eq:nMF_equi}.
On the methodological side much work has been done on applying
inverse Ising techniques to
synchronous versions of Ising dynamics~\cite{RoudiJohn2011a,RoudiHertz11DynamicTAP,mezard2011exact,zhang2012inference};
this work will not be covered here.
Dynamic mean-field inference as used below was originally developed
for synchronous updates in~\cite{KappenSpanjers2000},
see also~\cite{AurellMahmoudi2012}.
Inference in more realistic (and more complex) models from neuroscience
has also been carried out, but is beyond the scope of this review,
\textit{see} \cite{pillow2008spatio,cocco2009neuronal,mastromatteo2011criticality}.

\subsection{Mean-field Inference}
\label{sec:asynchronous-Ising-models-mean-field}
We now derive versions of \nMF\ and \TAP\ inference for asynchronously updated kinetic
models following~\cite{Zeng2011}.

For kinetic Ising model with Glauber dynamics, the state of spin $i$ is time dependent $s_i(t)$, thus the
time-dependent means and correlations are naturally defined as
\begin{equation} \label{eq:M-C}
\begin{aligned}
  m_i(t) &= \left<s_i(t)\right> \\
  c_{ij}(t_0+\tau,t_0) &= \left<s_i(\tau+t_0)s_j(t_0)\right> -m_i(\tau+t_0) m_j(t_0).
\end{aligned}
\end{equation}
Then, with the master equation (\ref{master_equation}) and the flipping rate (\ref{flipping_rate}), we have
equations of motion for means and correlations as
\begin{equation}  \label{equation_of_motion-m}
 \frac{dm_i(t)}{dt} = -m_i(t) + \langle\tanh\left[H_i(t)\right]\rangle.
\end{equation}
\begin{equation} \label{equation_of_motion}
 \frac{d\langle s_i(t)s_j(t_0)\rangle}{dt} =  - \langle s_i(t)s_j(t_0)\rangle + \langle \tanh\left[H_i(t)s_j(t_0)\right]\rangle.
\end{equation}
In the forward problem of statistical physics we would here have the \textit{closure problem}:
the left-hand side is the time derivative of an average while the right-hand side contains
terms of an average of a higher order.
In the inverse problem we start by observing that
the term on the
left-hand side and the first term on the right-hand side of equation \eqref{equation_of_motion-m}
and \eqref{equation_of_motion}
can be
taken from data. The second term on the right-hand side contains
averages of the $\tanh$ function and involves all kinds of higher-order correlations.
The equations thus have to be closed with respect to these terms, but
in a slightly different way in the forward problem.

We introduce the notation
\begin{equation}
b_i = \theta_i + \sum_j J_{ij}m_j
\end{equation}
for the non-fluctuating part of the argument of the $\tanh$
and rewrite $H_i(t) = \theta_i + \sum_j J_{ij}s_j(t)$ as
\begin{equation}
H_i \equiv b_i + \sum_j J_{ij} \delta s_j(t)
\end{equation}
where the sum depends on the
fluctuating term
$\delta s_i(t) = s_i(t)-m_i $.
In lowest order we neglect fluctuations in altogether and close the equation for magnetizations as
\begin{equation}  \label{equation_of_motion-m-closure}
 \frac{dm_i(t)}{dt} = -m_i(t) + \tanh b_i(t)\quad\hbox{(Lowest order closure)}
\end{equation}
If this equation reaches a stationary state it must satisfy $m_i = \tanh b_i$,
which we recognize as the equation of physical mean-field, \eqref{eq:mean-field}.
To the same lowest order \eqref{equation_of_motion} is
$\frac{d\langle s_i(t)s_j(t_0)\rangle}{dt} =  - \langle s_i(t)s_j(t_0)\rangle + m_i(t)m_j(t)$
which relaxes to the uncorrelated state.


The first non-trivial equation is obtained by expanding
\eqref{equation_of_motion} to first order which gives
\begin{equation}\label{expan_H}
  \langle s_i(t)s_j(t_0)\rangle + \frac{d\langle s_i(t)s_j(t_0)\rangle}{dt} = m_im_j + (1-m_i^2)\left(\sum_jJ_{ik}\langle \delta s_k(t)\delta s_j(t_0)\rangle\right)
\end{equation}
where we have used \eqref{equation_of_motion-m-closure} and stationarity to
identify the derivative of the $\tanh$ function as $(1-m_i^2)$.
Introducing
\begin{equation}\label{C_t_t_0}
  C_{ij}(t,t_0) = \langle\delta s_i(t)\delta s_j(t_0)\rangle = \langle s_i(t)s_j(t_0)\rangle -m_im_j.
\end{equation}
and
\begin{equation}\label{D}
\begin{split}
  D_{ij}(t,t_0) &= C_{ij}(t,t_0) + \frac{dC_{ij}(t,t_0)}{dt}\\
\end{split}
\end{equation}
we have
\begin{equation}
   \label{eq:D-definition}
   D_{ij}(t,t_0) = (1-m_i^2)\sum_kJ_{ik}C_{kj}(t,t_0)
\end{equation}
While this equation holds (to this order) for any two times $t$ and $t_0$
it is especially convenient in the limit
$t\rightarrow t_0$. Similarly to the procedure in
naive mean-field inference
\eqref{eq:nMF_equi} we can then invert \eqref{eq:D-definition} to arrive
at an \textit{asynchronous mean field inference} formula
\begin{equation}  \label{eq:J_nMF_asyn}
  J^{*,asyn-nMF} = A^{-1}DC^{-1},
\end{equation}
where $A$ is the diagonal matrix given by $A_{ij} = \delta_{ij}(1-m_i^2)$.
Equation (\ref{eq:J_nMF_asyn}) is a linear matrix equation with respect to $J_{ij}$. We can solve it for $J_{ij}$ directly for asynchronous Ising models.

 \begin{figure}[ht]
 \includegraphics[width=0.9\textwidth]{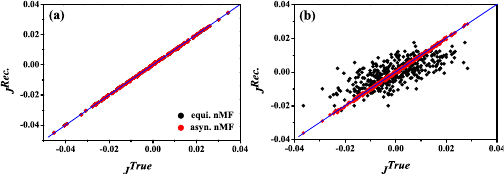}
 \caption{ The scatter plots for the true tested couplings versus the reconstructed ones. (a) reconstruction for the symmetric SK model with $k=0$; (b) inference for the asymmetric SK model with $k=1$. Red dots, inferred couplings with asynchronous  \nMF\ approximation; black dots, inferred ones with equilibrium  \nMF\ approximation.
 The recovered asynchronous $J_{ij}$s in (a) are symmetrized while no symmetrization for them in (b).
 The other parameters for both panels are $g=0.3$, $N=20$, $\theta=0$, $L=20\times 10^7$.
 } \label{fig:Equi-kinetic-nMF}
 \end{figure}

Figure \ref{fig:Equi-kinetic-nMF} shows the scatter plots for the tested couplings versus the recovered ones.
The tested model for Figure \ref{fig:Equi-kinetic-nMF}(a) is the symmetric SK model with $k=0$ in equation (\ref{Jij}) while fully asymmetric SK with $k=1$ for figure \ref{fig:Equi-kinetic-nMF}(b).
The couplings are reconstructed by the equilibrium \nMF\ (\ref{eq:nMF_equi}) (black dots) and the asynchronous \nMF\ (\ref{eq:J_nMF_asyn}) method (red dots) respectively.
As shown in figure \ref{fig:Equi-kinetic-nMF}(a), both methods have the same ability to recover the tested symmetric SK model. Here, the data length $L=20\times 10^7$. Nevertheless, the couplings inferred by the asynchronous \nMF\ needs to be symmetrized to  keep the same results with that from equilibrium \nMF, especially for short data length (not shown here).  Figure \ref{fig:Equi-kinetic-nMF}(b) shows that,  for the fully asymmetric SK model with $k=1$, the asynchronous \nMF\ works much better than the equilibrium \nMF. This clearly shows that equilibrium inference methods are typically not suitable for non-equilibrium processes, while asynchronous inference works for both equilibrium and non-equilibrium process.

By a similar procedure we can also derive a higher-order approximation,
which we refer to as \textit{dynamic TAP}.
The starting point is to redefine the $b_i(t)$ term in the $\tanh$
to include a term analogous to the static \TAP\ equation \eqref{eq:TAP}.
We then first have
\begin{equation}\label{T_i}
  H_i(t) = b_i- m_i\sum_{k\neq i}J_{ik}^2(1-m_k^2) + \sum_{k}J_{ik}\delta s_k(t).
\end{equation}
From which the lowest-order equation for the stationary state
is of the \TAP\ form.
The second step is to expand the $\tanh$ function in (\ref{equation_of_motion}) around
$b_i-m_i\sum_{k\neq i}J_{ik}^2(1-m_k^2)$
to the third order and to keep terms up to third order in $J$.
In this way
we get an inference
formula, which is formally the same as in the nMF approximation,
\begin{equation}\label{J_TAP}
J^{*,asyn-TAP}= A^{-1} D C^{-1}.
\end{equation}
where only the matrix $\textbf{A}$ is different
\begin{equation}\label{A_TAP}
A_{ij}=\delta_{ij}(1-m_i^2)\left[1-(1-m_i^2)\sum_j
J_{ij}^2(1-m_j^2)\right].
\end{equation}
Equation (\ref{J_TAP}) is a function of the couplings $\textbf{J}$, and therefore
it is a nonlinear equation for matrix $\textbf{J}$.

Equation \eqref{J_TAP} could be solved for $\textbf{J}$ though two approaches. One iterative
way is starting from reasonable initial
values $J_{ij}^0$, and inserting them in the RHS of formula (\ref{J_TAP}).
The resulting $J_{ij}^1$ is the solution after one iteration.
They can be again replaced in the RHS to get the second iteration results and so on.
\begin{equation}
J^{t+1} = A(J^t)^{-1}DC^{-1}
\end{equation}
An alternative way is solving it by casting the
inference formula to a set of cubic equations. For equation (\ref{A_TAP}),
denoting
\begin{equation}\label{F}
F_i=(1-m_i^2)\sum_{j}J_{ij}^2(1-m_j^2)
\end{equation}
and plugging it into equation (\ref{J_TAP}), and then we get the following equation
for $J_{ij}$:
\begin{eqnarray}\label{J_nMF_TAP}
J_{ij}^{asyn-TAP}= \frac{V_{ij}}{(1-m_i^2)(1-F_i)}
\end{eqnarray}
where $V_{ij}=[DC^{-1}]_{ij}$. Substituting equation (\ref{J_nMF_TAP}) with that in equation (\ref{F}), we
obtain the cubic equation for $F_{i}$ as
\begin{equation}\label{F_TAP}
F_i(1-F_i)^2-\frac{\sum_{j}V_{ij}^2(1-m_j^2)}{1-m_i^2}=0.
\end{equation}
With the obtained physical solution for $F_i$, we get the reconstructed couplings $J^\texttt{TAP}$ as
\begin{equation}\label{F_TAP_nMF}
 J_{ij}^{asyn-TAP}=\frac{J_{ij}^{asyn-nMF}}{1-F_i}.
\end{equation}

\subsection{Maximum-likelihood Inference}
\label{sec:asynchronous-Ising-models-ML}
To emphasize how different is inference from a time series compared
to from samples, we will now show that maximum likelihood inference
of such dynamics from such data is possible. We will also show
that this approach admits approximation schemes different
from mean-field. The presentation will follow \cite{Zeng2013}.

The log-likelihood of observing a full time series of a set
of interacting spins is analogous to the probability of a history of
a Poisson point process~\cite{VanKampen}.
The probability space of events in some time period $[0:t]$
consists of the number of jumps ($n$), the times of
these jumps ($t_1,t_2,\ldots,t_n$)
and which spin jumps
at each time ($i_1,i_2,\ldots,i_n$). The measure over this space
is proportional to the uniform measure over $n$ times a weight
$$\mu_{i_1}^{(1)}dt_1 \cdots \mu_{i_n}^{(n)}dt_n \cdot \exp\left(-\mu^{(1)}t_1-\mu^{(2)}(t_2-t_1)-\cdots -\mu^{(n+1)}(t-t_n)\right)$$
where $\mu_{i_n}^{(n)}$ is the jump rate in open time interval $(t_{i-1}:t_i)$ of the event
that actually took place at time $t_i$, and $\mu^{(n)}=\sum_i \mu_{i}^{(n)}$.
We recall from the discussion of
the Gillespie algorithm that in the open time interval $(t_{i-1}:t_i)$
all the rates stay the same, and that the length of the interval is
an exponentially distributed random variable with parameter which is the sum of all the rates.
In another time interval some or all of the rates can be different.

A rigorous construction of
the above path probability can be found in Appendix A of~\cite{KipnisLandim1999}.
Here we will follow a more heuristic approach and introduce a
small finite time $\delta t$ such that we can use the first simulation
approach discussed above in Section~\ref{sec:kinetic-Ising}.
The objective function to maximize is then
\begin{equation} \label{LL2}
\calL=\sum_{i,t}\log \left[(1-\gamma \delta t)\delta_{s_i(t+\delta t),s_i(t)}+\gamma \delta t \frac{{e}^{s_i(t+\delta t)H_i(t)}}{2\cosh H_i(t)}\right].
\end{equation}
The sums in (\ref{LL2}) go over all spins $i$ and all times separated by the small
increment $\delta t$.
The terms in (\ref{LL2}) can be understood as the lowest order approximation (linear in $\delta t$)
of $\log \prod_{it} P_i(s_i(t+\delta t) |\mathbf{s}(t))$
where $P_i$ is the conditional probability of spin $i$ at time $t+\delta t$,
conditioned on the configuration of all spins at time $t$.
Maximum likelihood inference of dynamics from a time series is
therefore analogous to pseudo-maximum likelihood \eqref{PML} from independent samples.
At the price of potentially very many and very biased samples (at most times no spin will jump)
this points to that inference from a time series is a fundamentally easier task.

Separating times with and without spin flips \eqref{LL2},  the resulting learning rules will be
\begin{equation} \label{eq:der}
\begin{split}
\delta J_{ij} &\propto \frac{\partial \calL}{\partial J_{ij}}\\
 &= \sum_{\emph{flips}} [s_i(t+\delta t)-\tanh(H_i(t))]s_j(t)
+ \frac{\gamma \delta t}{2}\sum_{\emph{no} \hspace{2 pt} \emph{flips}} q_i(t) s_i(t+\delta t)s_j(t),
\end{split}
\end{equation}
with $q_i(t)\equiv [1-\tanh^{2}(H_i(t))]$, and it includes the rule for the $\theta_i$ with the convention $J_{i0} = \theta_i$, $s_0(t) = 1$.
Following \cite{Zeng2013} where we also considered the case that the times where nothing happens are known,
we will refer (\ref{eq:der}) as the ``spin-history-only'' (``\textbf{SHO}'') algorithm.

Similarly to mean-field inference \eqref{eq:der} can also be averaged
which gives the learning rule
\begin{equation}
\delta J_{ij}\propto \gamma^{-1}{\dot C}_{ij}(0)+C_{ij}(0) -\langle \tanh(H_i(t))s_j(t)\rangle. \label{eq:J2av}
\end{equation}
which we refer to as \textbf{AVE}~\cite{Zeng2013}.
\textbf{AVE} requires knowing equal-time correlations, their derivatives at $t=0$, and $\langle \tanh(H_i(t))s_j(t)\rangle$.
This latter quantity depends on the model parameters (through $H_i(t)$), so, in practice,
estimating it at each learning step requires knowing the entire spin history, the same data as needs \SHO\ learning.

All of four methods now introduced to infer parameters from a time series (\nMF, \TAP, \SHO\ and \AVE) will produce a fully
connected network structure. Similarly to inverse Ising from samples we may want to include
$L_1$ penalties to get the graphical structure \cite{wainwright2007high}.
Such effects are considered in \cite{Zeng2014L1}, showing
that inferring the sparsity structure from time series data is both a feasible and reliable procedure.

\subsection{Performance of kinetic Ising inference methods}
In this section, performance tests of the four
above introduced algorithms for recovering parameters in asynchronous Ising models are presented.
We compared the performance of two ML algorithms \SHO, and \AVE\ to each other and to two mean-field algorithms \nMF\ and \TAP.

The tested model is as discussed above the fully asymmetric SK model ($J_{ij}$ is independent of $J_{ji}$), $J_{ij}$s  are identically and independently distributed Gaussian variables with zero means and variance $g^2/N$.
As a performance measure, we use the mean square error ($\epsilon$) which measures the
$L_2$ distance between the inferred parameters and the underlying parameters used to generate the data
\begin{equation}\label{MSE}
  \epsilon = \frac{\sum_{i\neq j}(J_{ij}^{*}-J_{ij}^{True})^2}{N(N-1)}.
\end{equation}
where $J_{ij}^{True}$ are the true values of interactions and
$J_{ij}^{*}$ are the inferred ones.
We study the reconstruction error for different data length $L$, system size $N$, external field $\theta$ and coupling strength $g$.

\begin{figure}[ht]
\centering
\includegraphics[width=0.9\textwidth]{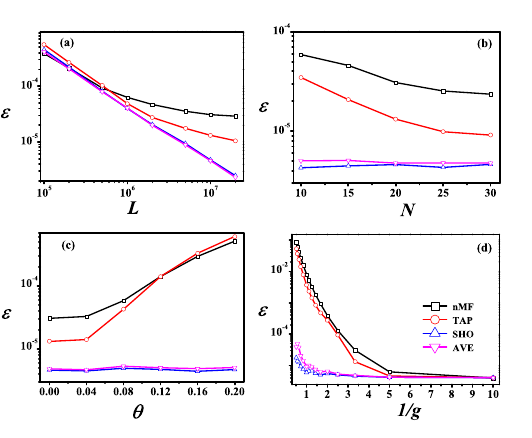}\\[5pt]  
\caption{Mean square error ($\epsilon$) versus (a) data length $L$, (b) system size $N$, (c) external field $\theta$ and (d) temperature $1/g$. Black squares show nMF, red circles, TAP, blue up triangle SHO and pink down triangle AVE respectively. The parameters are $g=0.3$, $N=20$, $\theta=0$, $L=10^7$ except when varied in a panel.}
\label{fig:MSE}
\end{figure}

Figure \ref{fig:MSE} shows the performance of these algorithms.   Each panel also shows two ML-based learning methods \SHO\ and \AVE\ appear to perform equally well for large enough $L$ since they effectively use the same data (the spin history). Note however the opposite trend in figure~\ref{fig:MSE}(a) shows the reconstruction getting better with longer data length $L$ for both ML and mean-field based methods.
Figure~\ref{fig:MSE}(b) shows that the MSE for the ML algorithms is
insensitive to $N$, while two mean-field algorithms improve as $N$
becomes larger; in these calculations, the average numbers of
updates and flips per spin were kept constant, taking $L = 5\times10^5N$).
Figure~\ref{fig:MSE}(c) shows that the performance of two ML algorithms is also not sensitive at all to $\theta$, while nMF and TAP work noticeably less well with a non-zero $\theta$. The effects of (inverse)$g$ are depicted in figure~\ref{fig:MSE}(d). For fixed
$L$, all the algorithms do worse at strong couplings (large $g$). The nMF and TAP do so in a much more clear fashion at smaller $g$, growing approximately exponentially with $g$ for $g$ greater than $\approx 0.2$.   In the weak-coupling limit, all algorithms perform roughly similarly, as already seen in figure~\ref{fig:MSE}(a).

To summarize, the ML methods recover the model better, but in general more slowly. The  mean-field based learning rules (\nMF\ and \TAP) are much faster in inferring the couplings but have worse accuracy compared with that of the ML-based iterative learning rules (\AVE, \SHO).


\section{Example Applications of Asynchronous Ising Model}
\label{sec:applications-Ising}
Inverse Ising problems have been applied to a wide rage of
data analysis, ranging from equilibrium reconstruction methods to kinetic ones.
In this section, based on~\cite{Zeng2013} and \cite{zeng2014financial},
we will present as illustrations applications to one data set
of neuronal spike trains, and one data set on transaction data of stocks on financial market.
Both areas have been investigated extensively in the last ten years.
We refer to \cite{ROUDI2015, COCCO2017103,huang2017theory,Massobrio2016,Pillow2019inferring,Berry2020salamanderData,hoang2019network} for more recent neuronal data
and and discussions of inference in this context
and to~\cite{Bacry2015,ma2015dynamics,borysov2015,Zarinelli2015,li2016network,fan2017entropies,zhao2018stock,becker2018maximum,alossaimy2019using,Bucci2019,Hoffmanneaav2020, ikeda2020reconstruction} a for a sample of
contributions considering financial data.

For the neuronal data, we show two ML-based learning rules.
When considering the data as a time series we use the
\AVE\ method \eqref{eq:J2av}, while when considering the same data as
independent samples from the Gibbs distribution \eqref{Gibbs-distribution}
we use Boltzmann machine (\textbf{BM}) (introduced below).
We find that for this data the couplings between the neurons obtained are comparable.
This means that although there is no a priori for this to be so,
the dynamic process of this neuron system apparently satisfies detailed
balance or has a stationary distribution of the form of \eqref{Gibbs-distribution}
for other reasons.
One clear difference is the self-couplings from one neuron to itself which are absent in
\eqref{Gibbs-distribution} but which are typically present in the dynamic model.
A further difference is that to infer parameters from \eqref{Gibbs-distribution}
using samples, those samples have to be generated by Monte Carlo procedure.
Although both methods are based on ML, the dynamic version is thus considerably
faster than \textbf{BM}.

For the financial stock trades data we show two mean-field-based algorithms.
When considering the data as a time series we use the
asynchronous \nMF\  method of \eqref{eq:J_nMF_asyn} while when considering the same data as
independent samples from \eqref{Gibbs-distribution},
we use naive mean-field inference (here equilibrium \nMF) of \eqref{eq:nMF_equi}.
We note that we here apply inverse Ising inference to
binary data obtained by transforming a time series of financial transactions (see below).
Again we find that the results from the two procedures are comparable,
except that asynchronous \nMF\ allows the inference of self-couplings, as well as directed links (asymmetric couplings).

\subsection{Case 1: Reconstruction of a neuron network from spiking trains}
Neurons are the computational units of the brain.
While each neuron is actually a cell with complicated internal
structure, there is a long history of considering simplified
models where the state of a neuron at a given time is
a Boolean variable. Zero (or down, or -1) then means resting,
and one (or up, or +1) means firing, or having a spike of activity.
In most neural data most neurons are resting most of the time.

\emph{Data description and representation of data.}
The neuronal spike trains are from salamander retina under stimulation by a repeated 26.5-second movie clip.
This data set records the spiking times for neurons and has a data length of 3180 seconds (120 repetitions of the movie clip).
Here, only the first $N = 20$ neurons with highest firing rates in the data set are considered.     The data has been binned with time windows of 20 ms (the typical time scale of the auto-correlation function of a neuron) in the previous study \cite{Berry2006functional}.
However, since we are using the kinetic model, we could study this data set using a much shorter time bin which leads  low enough firing rates and (almost) never more than one spike per bin.
Then, the temporal correlations with time delays between neuron pairs as well as the self-correlations become important.

For the asynchronous Ising model, the time bins are $\delta t = 1/(\gamma N)$.   For neuronal data, $\gamma$ can be interpreted as the inverse of the time length of the auto-correlation function which is typically 10 ms or more \cite{Berry2006functional}.
To generate the binary spin history from this spike train data set,  the spike trains should be separated into time bins with length $\gamma\delta t = 1/20$. This means the size of time bins should be chosen as $\delta t = 1/(20\gamma) = 0.5$ ms.
The spin trains can be transformed in to binaries as follows: a $+1$ is assigned to every time bin in which there is a spike and a $-1$ when there is no spikes.
To avoid the case that
the translation always end up with isolated instances of $+1$ and superfluous $-1$s, the  memory process for each neuron is introduced to the data set. It is a time period with an exponential distribution with mean of $1/\gamma$ in the data translation. Denote the total firing number of neuron $i$ as $F_i$, and $t_i^f$ as the firing time of $f$th spike for neuron $i$, where $i=1,...,N$ and $f=1,...,F_i-1$, then the mapping of the spike history is follows:
\begin{equation}\label{Si-neuron}
  \mathrm s_i(t) = \left\{ \begin{array}{ll}
   1, & \text{if}~~t\in \bigl[t_i^f,_{\min}(t_i^{f+1},t_i^n+X)\bigr)~~\text{with}~~X\sim \exp(\gamma^{-1})   \\
   -1, & \text{otherwise} \end{array} \right.
\end{equation}
where $X$ is a period drawn from exponential distribution with mean 10 ms. By this way, we obtain the asynchronous type of data that are needed for the asynchronous model.

\emph{Inference methods.}
For this fairly small system we use two types of ML to learn the parameters of \eqref{Gibbs-distribution}.
In the equilibrium case \eqref{ML}
this can be done with the iterative method called Boltzmann machine (\BM)
which is defined as follows:
\begin{equation*}
  \begin{aligned}
    \delta \theta_i &= \eta\left(\langle s_i\rangle_{Data} - \langle s_i\rangle_{Model}\right),\\
    \delta J_{ij} &= \eta\left(\langle s_is_j\rangle_{Data} - \langle s_is_j\rangle_{Model}\right).
  \end{aligned}
\end{equation*}
In above $\eta$ ``learning rate'' is a relaxation parameter.
For larger systems \BM\ does not scale since computing the ensemble averages
$\langle s_i\rangle_{Model}$ and
$\langle s_is_j\rangle_{Model}$ is costly, but for the data under consideration here it is a feasible method.
When retaining the time series nature of the data we on the other hand use the
\AVE\ learning rule of equation \eqref{eq:J2av}.

\begin{figure}[ht]
\centering
\includegraphics[width=0.5\textwidth]{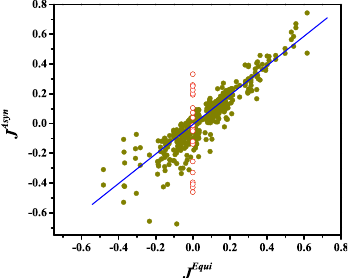}
\caption{Inferred asynchronous versus equilibrium couplings for retinal data.
Red open dots show the self-couplings which by convention are equal to zero for the equilibrium model.}
\label{asyn_equ}
\end{figure}

\emph{Inference Results. }
In the current inference of retina functional connections, the value of model parameters like window size $\delta t$, inverse time scale $\gamma$ are set as \emph{a priori} according to the previous studies on equilibrium Ising model. This avoids systematic studies over the value of parameters.

As presented in figure \ref{asyn_equ}, the inferred couplings by \BM\ and asynchronous kinetic Ising model are very close to each other.
We also tested what
happens to the couplings of the asynchronous model if during learning we symmetrized the couplings matrix at each iteration by adding its transpose to itself and dividing
by two and also putting the self-couplings to zero. We find that the resulting asynchronous couplings get even closer to the equilibrium ones, which is consistent to the conclusion for kinetic Ising data.

However, the asynchronous model allows the inference of self-couplings (diagonal elements of the coupling matrix) which are not present in the equilibrium model. As shown in figure \ref{asyn_equ}, the diagonals from the equilibrium model equals to zeros by convention and denoted by the open red dots. Furthermore, to be different from the symmetric couplings by the equilibrium model, the asynchronous model provides more details as the inferred couplings are directed and asymmetric.

This result provides a guide for the use of the equilibrium Ising model: if the asynchronous couplings were far away form the equilibrium ones, it would imply that the real dynamical process did not satisfy the Gibbs equilibrium conditions and that the final distribution of states is not the Gibbs equilibrium Ising model.  Since inferring the equilibrium model is an exponentially difficult problem, requiring time consuming for Monte Carlo sampling while the asynchronous approach does not. The asynchronous learning rules thus allow the inference of functional connections that for the retinal data largely agree with the equilibrium model, but the inference is much faster.

\subsection{Case 2: Reconstruction of a finance network}

In this case study, we present equilibrium \nMF\ (\ref{eq:nMF_equi}) and asynchronous \nMF\ (\ref{eq:J_nMF_asyn}) algorithm to infer a financial network from trade data with 100 stocks.       The recorded time series are transformed into binaries by local averaging and thresholding. This introduces additional parameters that have to be studied extensively to understand the behavior of the system.
The inferred couplings from asynchronous \nMF\ method is quite similar to the equilibrium ones.
Both produce network communities have similar industrial features. However, the asynchronous method is more detailed as they are directed compared with that from the equilibrium ones.

\emph{Data description and representation.}
The data was transactions recordings on the New York Stock Exchange (NYSE) over a few years.
Each trade is characterized by a time, a traded volume, and a price.
We only focus on the trades for 100 trading days between 02.01.2003 and 30.05.2003.
However, trading volume and trading time only are utilized in the study. To avoid the opening and closing periods of the stock exchange, $10^4$ central seconds of each day are employed as in \cite{mastromatteo_marsiliJSTAT2011}.
Two parameters are introduced to the data transform as the sliding window is adopted. One is the size of the sliding time window (denoted as $\Delta t$), the other one is the shifting constant which is fixed as $1$ second.

For stock $i$, the sum of the volumes $V_i(t,\Delta t)$ traded in window $[t, t+\Delta t)$, is compared with a given volume threshold $V^{th}_i = \chi V^{av}_i\Delta t$, where $V^{av}_i$ is the average (over the whole time series) volume of the considered stock traded per second, and $\chi$ a parameter controlling our volume threshold:
\begin{equation}\label{s_mapping}
     \mathrm s_i(t) = \left\{
     \begin{array}{ll}
   1, & if~~V_i(t, \Delta t) \ge V^i_{th}  \\
   -1, & if~~V_i(t, \Delta t) < V^i_{th} \end{array} \right.
\end{equation}
We explored the parameters $\Delta t$ and $\chi$   systematically for the inference with the goal that to find values of the parameters which yield inferred couplings containing interesting information.

\begin{figure}[ht]
\centering
\includegraphics[width=0.5\textwidth]{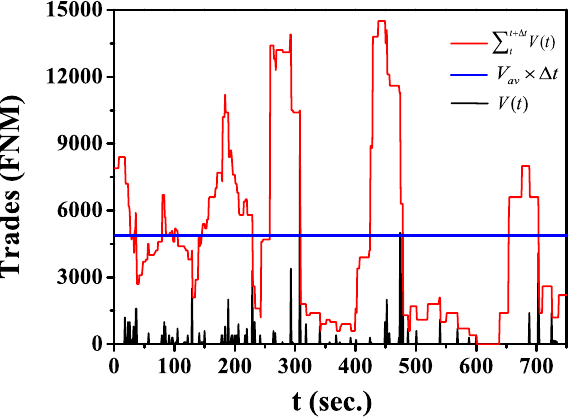}
\caption{Traded volume data for the stock of Fannie Mae (FNM), a mortgage company. Black line for time series of traded volumes $V_i(t)$, red for summed volumes during time interval $\Delta t$, blue for the threshold $V_i^{th}=\chi V_i^{av}\times \Delta t$. Parameters: $\Delta t = 50 s$ and $\chi=1$.}
\label{fig:finance-timeseries}
\end{figure}

Figure \ref{fig:finance-timeseries} shows the traded volume information for a mortgage company Fannie Mae(FNM). With the mapping approach described in equation (\ref{s_mapping}), we have $+1$s above the blue threshold line in figure \ref{fig:finance-timeseries} while $-1$s below that line. Then the asynchronous data is ready for the network reconstruction.

\emph{Inference methods.}
With the transformed binaries, the magnetization $m_i$ and correlations $C_{ij}(\tau)$ are defined as (\ref{eq:M-C}).
With them, two different inference methods with nMF approximation are utilized for the reconstruction.
\begin{itemize}
  \item Equilibrium \nMF\ ($i\ne j$), which only focuses on equal time correlations \cite{Kappen1998}
  $$J_{ij} =  - C(0)_{ij}^{-1}$$
  \item Asynchronous \nMF\  \cite{Zeng2011},  uses the derivative of the time-lagged correlations $\dot C_{ij}(\tau)$, as shown in equation (\ref{eq:J_nMF_asyn}) and be rewritten as:
      $$J_{ij} = \frac{1}{1-m_i^2}\left(\frac{dC(\tau)}{d\tau}|_{\tau=0}C(0)^{-1}\right)_{ij}$$
\end{itemize}

\emph{Reconstruction Results.}
Massive explorations over different values of the window size $\Delta t$ and $\chi$s are complimented to achieve meaningful interactions between stocks.
A natural rough approach is to consider that couplings contain interesting information if they are big in absolute value: they indicate a strong interaction between stocks.
For asynchronous inference, the derivative of the time-lagged correlations $\dot{C}_{ij}(\tau)$ is computed through a linear fitting of this function $C_{ij}(\tau)$ using four points: $C(0)$, $C(\Delta t/5)$, $C(2\Delta t/5)$ and $C(3\Delta t/5)$.

\begin{figure}[ht]
    \centering
    \includegraphics[width=0.5\textwidth]{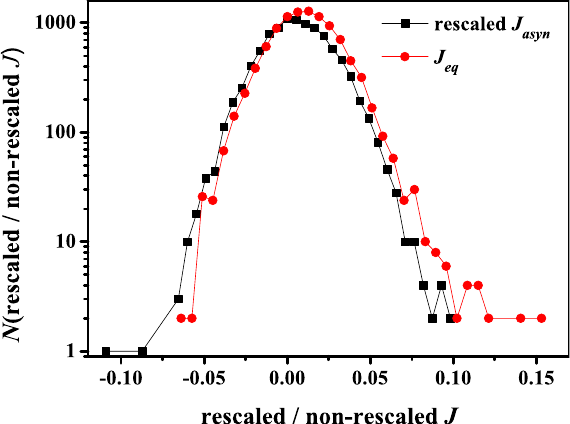}
    \caption{Histograms of inferred couplings by equilibrium \nMF\ and re-scaled asynchronous \nMF. Black squares for re-scaled $N(J_{asyn})$ to have the same standard deviation as $N(J_{eq})$. $\chi = 0.5$ and $\Delta t = 200$ seconds for both methods.}
\label{Fig:hist_Jeq_rescaled_Jasyn}
\end{figure}
\begin{figure}[ht]
    \centering
   \includegraphics[width=0.5\textwidth]{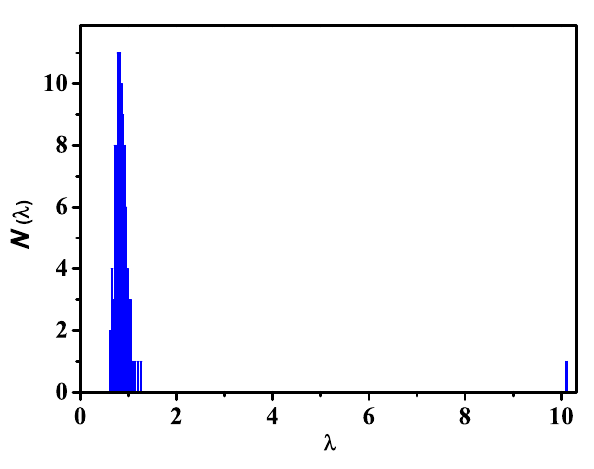}
   \caption{ Histograms of the eigenvalues of the equal time connected correlation matrix. Parameters: $\chi= 0.5$ and $\Delta t = 100$ seconds.}
\label{Fig:dist-eigenvalues}
\end{figure}

Figure \ref{Fig:hist_Jeq_rescaled_Jasyn} shows both inference methods give similar distributions of couplings. For comparison, the distributions are re-scaled so as to have the same standard deviation.
It can be remarked that the inferred couplings have a strictly positive mean and a long positive tail.
This prevalence of positive couplings can intuitively
be linked with the market mode phenomenon \cite{bury2013statistical, bouchaud2003theory, mantegna2003introduction, biely2008random}: a large eigenvalue appears, corresponding to a collective activity of all stocks, illustrated in figure \ref{Fig:dist-eigenvalues}.

The similarity of interaction matrices $J$ and $J'$ inferred from different methods can be measured by a similarity quantity $Q_{J,J'}$, which is defined as
\begin{equation}\label{Q}
  Q_{J,J'} = \frac{\sum_{i,j}J_{ij}J'_{ij}}{\sum_{i,j}\mathrm{max}(J_{ij}, J'_{ij})^2}
\end{equation}
This measurement compares elements of two matrices one by one and gives a global similarity measure. It takes real values between 1 (when $J_{ij} = J'_{ij}$ for all $i$ and $j$) and -1 ($J_{ij} = -J'_{ij}$ for all $i$ and $j$), and values close to zero indicate uncorrelated couplings.      The values of $Q$ is smaller than 0.02 in absolute value when all elements of the vectors $J_{ij}$ and $J'_{ij}$ are drawn independently at random from a same Gaussian distribution, of mean 0, and for different values of the standard deviation of this distribution. Here, the value of $Q$ for the inferred $J_{ij}$s by equilibrium \nMF\ and asynchronous \nMF\ is about $0.5$ with $\chi=0.5$ and $\Delta t = 50$ sec., which indicates these two methods are not independent to each other.

\begin{figure}[ht]
\centering
    \includegraphics[width=0.45\textwidth]{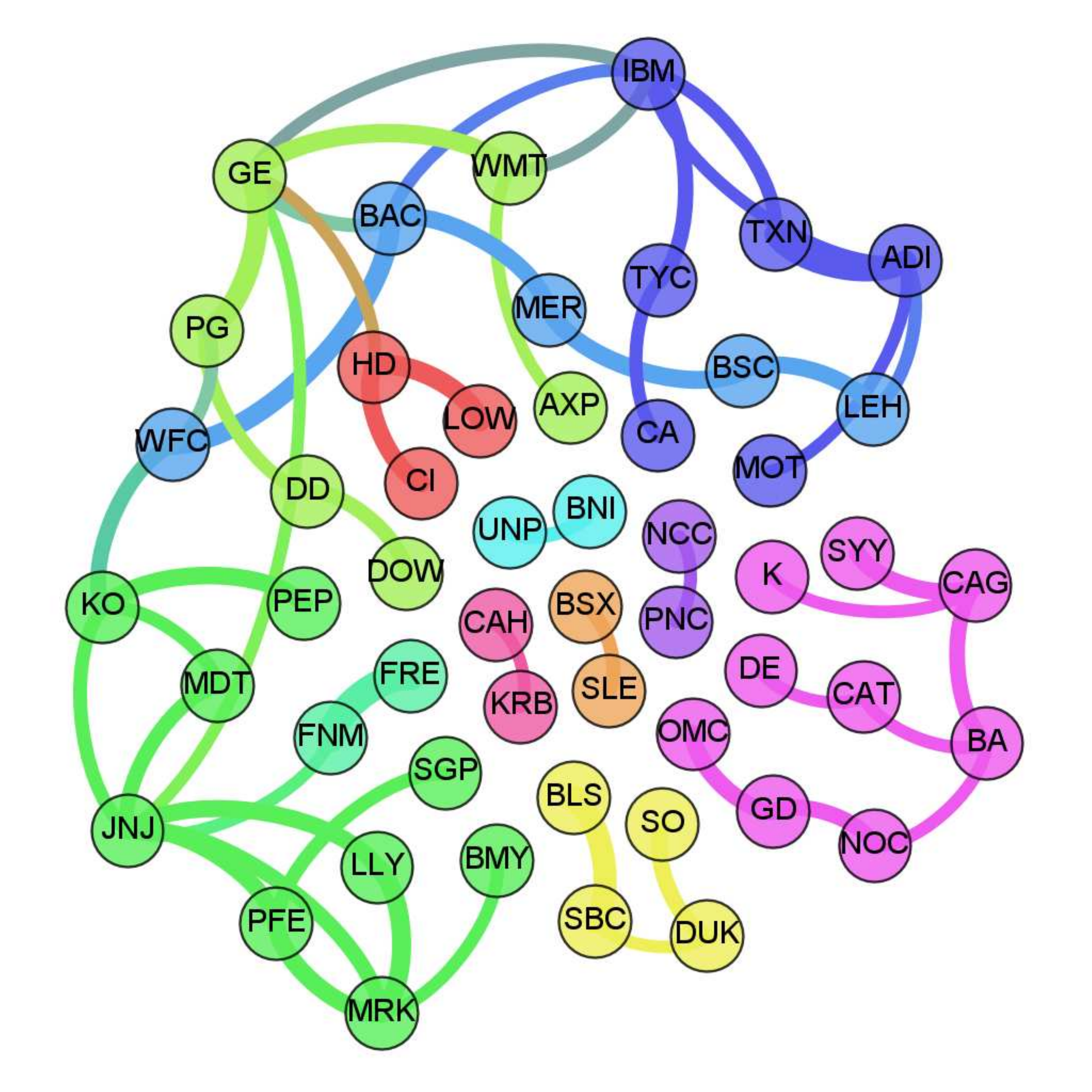}
    \includegraphics[width=0.45\textwidth]{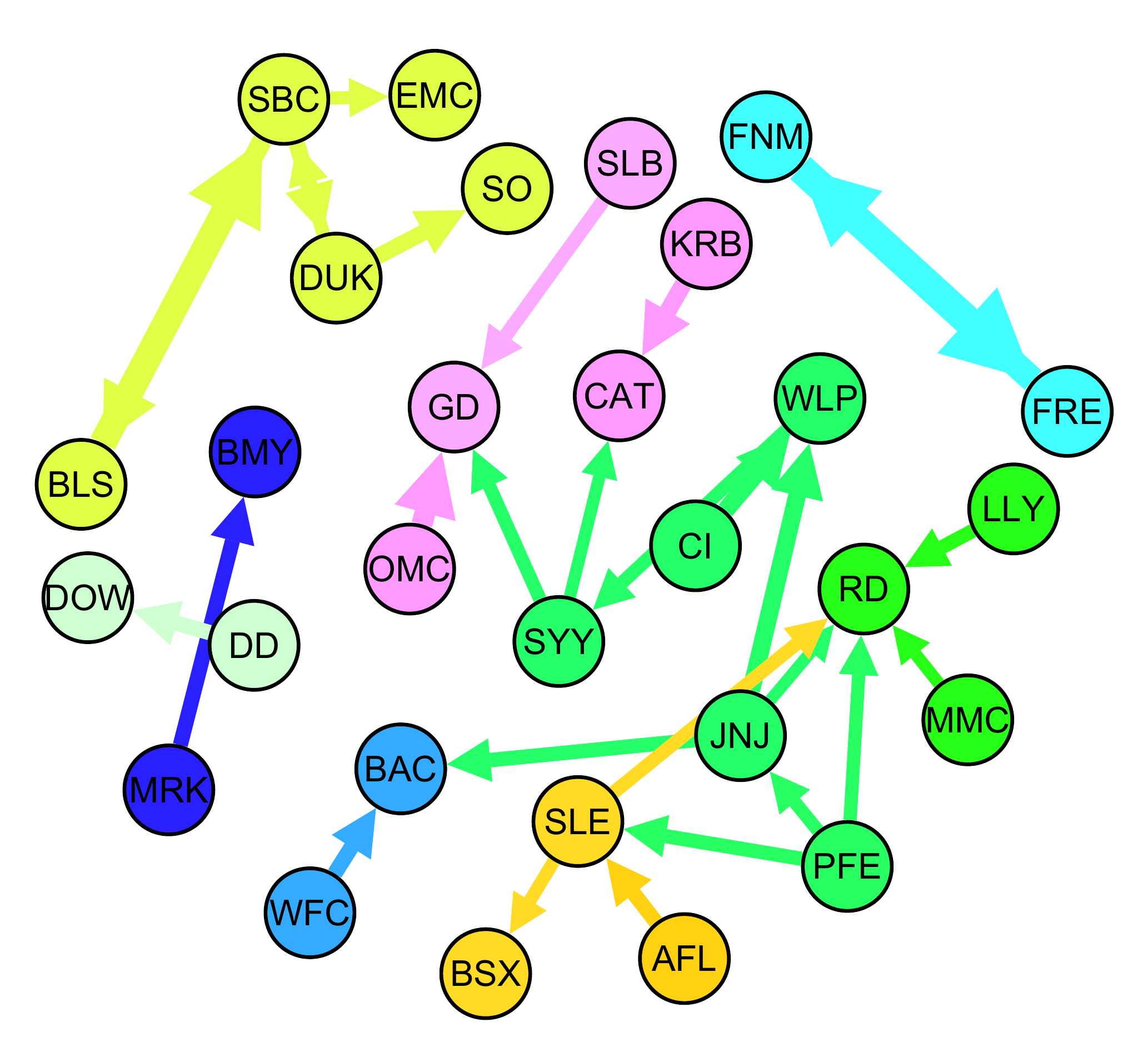}
    \put(-430,200){\textbf{(a)}}
    \put(-210,200){\textbf{(b)}}
    \caption{Inferred financial networks, showing only the largest interaction strengths (proportional to the width of links and arrows). Colors are indicative, and chosen
by a modularity-based community detection algorithm [16]. Parameters: $\chi = 0.5$
and $\Delta t = 100$ seconds.
(a): equilibrium inference; (b): asynchronous
inference, with $\tau = 20$ seconds.}
\label{fig:equi-asyn-network}
\end{figure}

Next, we will present two inferred financial networks that recovered by equilibrium and asynchronous \nMF\ method respectively.
As the inferred finance networks are densely connected, we focus only on the largest couplings, which can be explained by closely related activities of the considered stocks. Figure \ref{fig:equi-asyn-network}(a) shows that with equilibrium inference, more than half the stocks in the data can be displayed on a network where almost all links have simple economical interpretations.

The network of figure \ref{fig:equi-asyn-network}(a)  presents different communities, each color represents one industrial sector. They are mostly determined by a common industrial activity. Some of the links are very easy to explain with the proximity of activities (and often quite robust).  For instance,  the pairs FNM - FRE (Fannie Mae - Freddie Mac, active in home loan and mortgage), UNP - BNI (Union Pacific Corporation - Burlington Northern Santa Fe Corporation, railroads), BLS - SBC (BellSouth - SBC Communications, two telecommunications companies now merged in AT$\&$T),  DOW - DD (Dow - DuPont, chemical companies), MRK - PFE (Merck $\&$ Co. - Pfizer, pharmaceutical companies), KO - PEP (The Coca-Cola Company - PepsiCo, beverages).
These two last companies display strong links with the medical sector at different scales of volume and time, as KO here with MDT (Medtronic) and JNJ (Johnson $\&$ Johnson). This medical sector is itself linked to the pharmaceutical sector with PFE, MRK, LLY (Lilly), BMY (Bristol-Myers Squibb) and SGP (Schering-Plough). Telecommunications (BLS, SBC) are linked to electric power with DUK (Duke Energy),

GE (General Electrics) is for a large range of parameters a very central node, which is consistent with its diversified activities. Figure \ref{fig:equi-asyn-network}(a) presents the relation between PG (Procter $\&$ Gamble) and WMT (Walmart), both retailers of consumer goods, comes at this level of interaction strength through GE.

The banking sector as shown by a chain with light blue color is linked to the sector of electronic technology (with dark blue color).   Moreover, the defense and aerospace sector as shown in magenta is linked to engines and machinery with  (CAT) (Caterpillar Inc.) and DE (John Deere), and more strangely, to packaged food with CAG (ConAgra Foods), SYY (Sysco) and K (Kellogg Company).

Figure~\ref{fig:equi-asyn-network}(b) presents the results from asynchronous \nMF\ in the
same conditions. It shows that the results of equilibrium and asynchronous inference are consistent, and that asynchronous inference provides additional information, as it infers
an directed network. For instance,
the financial sector is directed and influenced by the medical sector also. The detailed descriptions for each stock can be found in \cite{zeng2014connectivity}.

From the network samples, we have the following two basic conclusions. First, they show market mode (most of the interaction strengths found are usually positive, which indicates that the financial market has a clear collective behavior) \cite{bouchaud2003theory, mantegna2003introduction} even only trade and volume information is considered. Stocks tend to be traded or not traded at the same time.

In addition, the strongest inferred interactions can be easily understood by similarities in the industrial activities of the considered stocks. This means that financial activity tends to concentrate on a certain activity sector at a certain time. For price dynamics this phenomenon is well-known \cite{bury2013statistical, biely2008random, kullmann2002time}, but it is perhaps more surprising that it appeared based also on only information of traded volumes.

\section{Fitness inference of population genetics}
\label{sec:population-genetics}
We now turn to inverse Ising (Potts) techniques applied to sequence-type biological data.
This has variously been called Direct Coupling Analysis (DCA)~\cite{Weigt-2009a,Morcos-2011a,Marks2011,Cocco-2018a}
and max-entropy modeling \cite{Tkacic2014,Stein-2015a};
as noted above, other names are also in use. We will use the terms inverse Ising and DCA
interchangeably.

A common feature of all these applications is that the input is a static table of $N\cdot L$ symbols.
Each row is a sequence of $L$ symbols from data, and there
are $N$ such rows ($N$ samples). A breakthrough application has
been to identify residues (amino acid molecules) that are spatially
close in proteins (chains of amino acids). The table then represents
a family of proteins with supposedly similar structure and supposedly same origin, and
each row is the amino acid sequence of a member of that family~\cite{Weigt-2009a,Morcos-2011a,Marks2011,Cocco-2018a}.
The basic idea is that two columns in the table (two positions in the protein structure)
have non-trivial statistical dependency if their joint variation
influence biological fitness. Such co-dependency in biological fitness is called \textit{epistasis}.
The most immediate cause of epistatsis among loci inside one gene coding for one protein
is through structure~\cite{EPISTASIS}.
Often this is pictorially motivated by a mutation changing
charge, hydrophilic/hydrophobic or size of one member of a residue pair,
which then changes the relative fitness of variants (alleles) of the other member of the pair.
In certain other cases dependencies discovered by DCA can be
attributed to other causes than structure \cite{Gueudre-2016a,Uguzzoni-2017a} but those cases appear to be relatively rare.

Many details are needed to turn the above to powerful tool in protein structure
prediction. One aspect is that proteins in a family typically
have different lengths, and that therefore the $N\cdot L$ table
is not directly taken from the data, but only after multiple sequence alignment, which
has to be done with the help of bionformatics software, or the ready alignment
taken from a data base such as PFAM~\cite{PFAM,PFAM-ref}.
Another is that predicting contacts is only one ingredient in a much larger
computational pipeline which uses inter-molecular force fields, predictions
on secondary structure and solvability and know-how developed in the
protein science community over many years. Still, impressive results have
been achieved~\cite{Ovchinnikov-2017a,Michel-2017b,Ovchinnikov-2018a}.
It should be noted that if the goal is to predict protein structure a purely data-driven approach is possible,
where a model of the deep neural network type is trained on large
training sets comprised of sequence-structure pairs.
As has been widely reported, such an approach from Google Deep Mind
currently outperforms model-based learning methods such as
DCA for this task~\cite{AlphaFold, AlphaFold-ref}. The price is computational cost beyond
what most academic researchers can afford, and lack interpretability of the inferred model,
which could be close to~\eqref{Gibbs-distribution}, but could also be very different.

Beyond protein structures DCA has been used to predict
nucleotide-nucleotide contacts of RNAs~\cite{DeLeonardis-2015a},
multiple-scale protein-protein interactions~\cite{Uguzzoni-2017a},
amino acid--nucleotide interaction in RNA-protein complexes~\cite{Weinreb-2016a},
interactions between HIV and the host immune system
\cite{Ferguson-2013a,Shekhar-2013a,Louie-2018a},
and other synergistic effects not necessarily related to spatial contacts~\cite{Figliuzzi-2016a,Hopf-2017a,CouceE9026}.
Of particular relevance for the following are applications of DCA
to whole-genome sequence data from bacterial populations,
in~\cite{Skwark-2017a} on \textit{Streptoccoccus pneumoniae}
and in~\cite{Schubert-2018a} on \textit{Neisseria gonorrhoeae}.
Standard versions of DCA are rather compute-intensive for genome-scale inference tasks,
but methodological speed-ups~\cite{Puranen-2017a,GaoZhouAurell2018}
and alternative approaches~\cite{Pensar2019} has been quickly developed.
Antibiotic resistance is an important medical problem throughout
the world, and so is the relative paucity of new drugs.
Combinatorial drug combinations are therefore promising
avenues to look for new treatment strategies. The
obstacle is the combinatorial explosion of combinations:
if there are $L$ potential individual targets there are $L^2$
potential target pairs, and so on.
The hope is that DCA could be one way (one out of many) to predict
which combinations may have an effect on the grounds that they are
already reflected as epistasis in natural sequence data.
In that respect it was promising that
Skwark et al in~\cite{Skwark-2017a}
were able to
retrieve interactions
between members of the Penicillin-Binding Protein (PBP) family
of proteins; resistance to antibiotics in the $\beta$-lactam family of compounds
is in \textit{S. pneumoniae} associated to alterations in their target enzymes, which are the PBPs~\cite{Hakenbeck2012}.

Evolution is a dynamic process.
We should imagine that the biological sequence data
used in DCA are as in Section~\ref{sec:time-series}
(or more involved).
The underlying dynamics is of $N_{tot}$ sequences
(a number which could change with time, but which we will assume constant)
of length $L$ (which could also change, but which we will also assume constant),
and which evolve evolving for a time $T$.
At the end of the process we sample $N$ sequences.
In protein data $T$ is typically of the order of hundreds of millions of years,
and the model is obviously simplified.
In the bacterial whole-genome data of~\cite{Skwark-2017a,Schubert-2018a}
$T$ may be as short as years or decades, and the model may be closer to reality.
In any case, the goal is
to infer fitness from the sampled sequences, and to understand when that
can (or cannot) reasonably be done by DCA.

We will structure the discussion as follows. In Section~\ref{sec:population-genetics-dynamics},
we will discuss dynamics of a population in a fitness landscape on which
there is a large literature both in population genetics and in statistical
physics. We will there define what we mean by fitness,
and introduce recombination.
In Section~\ref{sec:QLE-phase}, we will present
the important concept of Quasi-Linkage Equilibrium (QLE), originally due to Kimura,
and in Section~\ref{sec:interactions-fitness} we will state the relation
between inferred interactions and underlying fitness that hold in QLE.
Numerical examples and tests are presented in Section~\ref{sec:fitness-inference-synthetic}.

\subsection{Dynamics of a sexually reproducing population in a fitness landscape}
\label{sec:population-genetics-dynamics}
That there exists
formal similarities between the dynamics
of genomes in a population and entities (spins) in statistical physics
has been known for a long time.
Fokker-Planck equations to describe the change of probability distributions
over allele frequencies were introduced by Fisher almost a century ago~\cite{Fisher1922,Fisher-book},
and later, in a very clear a concise manner, by Kolmogorov~\cite{Kolmogorov1935}.
The link has been reviewed several times from the side of
statistical physics, for instance in~\cite{Peliti-1997} and~\cite{Blythe-2007a}.
Central to the discussion in the following will be
recombination (or sex), by which two parents give rise to an offspring
the genome of which is a mixture of the genome of the parents.
From the point of view of statistical physics recombination
is a kind of collision phenomenon. It therefore cannot be described
by linear equations (Fokker-Planck-like equations)
but can conceivably be described by nonlinear equations (Boltzmann-like equations).
The mechanisms to be discussed are of this type, where
Boltzmann's \textit{Stosszahlansatz} is used to factorize the collision operator.

All mammals reproduce sexually, as do almost all
birds, reptiles and fishes, and most insects.
Many plants and fungi can reproduce both sexually and asexually.
Recombination in bacteria is much less of a all-or-none affair.
Typically only some genetic material is passed
from a donor to a recipient, directly or indirectly.
The main forms of bacterial recombination are
conjugation (direct transfer of DNA from a donor to a recipient),
transformation (ability to take up DNA from the surroundings),
and transduction  (transfer of genetic material by the intermediary of viruses).
The relative rate of recombination in bacteria varies greatly
between species, and also within one species, depending on conditions.
As one example we quote a tabulation of the ratio of
recombination to mutation rate
in \textit{S. pneumoniae},
which has been measured to vary from less than one to over forty~\cite{Chaguza2016}.
There are long-standing theoretical arguments against the
possibility of complex life without sex, as a consequence
of Eigen's ``error catastrophe''~\cite{Eigen1971,Eigen2002}. It is likely that most forms
of life use some form of recombination,
albeit perhaps not all the time, and though the relative rate of
recombination to other processes may be small.
In the following we will eventually assume that
recombination is faster than other processes,
which may be as much the exception as the norm in bacteria and other microorganisms.
Such a ``dense-gas''  (using the analogy with collisions)
is however where there is an available theory
which can be used at the present time.

The driving forces of evolution are hence assumed to be genetic drift,
mutations, recombination, and fitness variations. The first refers to the element of chance; in a finite
population it is not certain which genotypes will reproduce and leave
descendants in later generations.
The last three describe the expected success or failure of different
genotypes.

Genetic drift can be explained by considering $N$ different genomes
$\mathbf{s}_1,\ldots,\mathbf{s}_N$. Under \textit{neutral evolution}
all genomes have equal chance to survive into the next generation,
but that does not mean all will do so. In a \textit{Wright-Fisher model}
one considers a new generation with $N$ new genomes
$\mathbf{s'}_1,\ldots,\mathbf{s'}_N$, where each one is
a drawn randomly with uniform probability from the previous generation.
The chance (or risk) that an individual does not survive from one generation to the next
is then $\left(1-\frac{1}{N}\right)^N$, which is about $e^{-1}\approx 37\%$.
Monte Carlo simulations of finite populations necessarily include
such effects where some successful individuals crowd out other less
fortunate ones.

Mutations are random genome changes described by mean rates. A model of
$N$ individuals evolving under mutations and genetic drift
which happen synchronously is also called a Wright-Fisher model, and
when they happen asynchronously a Moran model~\cite{Blythe-2007a}.
If the genome (or the variability of the genome) consists of only one biallelic locus
(one Ising spin, $L=1$) then the state of a population is given by the number
$k$ of individuals
where the allele is``up'' ($N-k$ individuals then have the ``down'' allele).
The dynamics of Moran model can be seen as the dynamics of $N$ spins
where each spin can flip on its own or can copy the state of another spin
(or do nothing).
It can also be seen as a transitions in a finite lattice
where $k=0$ means all spins are down, and $k=N$ means all spins are up.
The probability distribution over this variable $k$ changes by a Master equation
where the variable can take values $0,1,\ldots,N$.
If mutation rate is zero the two end states in the lattice
are absorbing: eventually all individuals will be up,
or all will be down. If mutation rate is non-zero but small,
the stationary probability distribution is centered on small and large
$k$ and transitions between the two macroscopic states happen only rarely. For a very pedagogical
discussion of these classical facts we refer to~\cite{Blythe-2007a}.

The evolution of the distribution over $L$ biallelic loci under mutations
has many similarities to \eqref{master_equation}, and always satisfies detailed balance \protect\footnotemark[7]{$^7~$}
\protect\footnotetext[7]{$^7~$ This is not generally true for more than
one allele per locus and a general mutation matrix.}.
As the rate $r_i(\s)$ in \eqref{detailed-balance-spin-flip-2}
the rate of mutations can and generally does depend on genomic position
(``mutation hotspots'') and on the alleles at other loci
(``genomic context'').
For theoretical discussion and simulations it is however
more convenient to assume an overall uniform flipping rate
as in \eqref{flipping_rate}.

A fitness landscape means a propensity
for a given genotype to propagate its genomic material to the next generation.
This propensity is a function of genotype, and called a fitness function.
It is important to note that this concept does not cover all that
fitness can mean in biology. Excluded effects are for instance
cyclical dominance where ``$A$ beats $B$'',  ``$B$ beats $C$''
and ``$C$ beats $A$''~\cite{MaynardSmith82,Nowak2004,ClaussenTraulsen2008,WangXuZhou2014,Liao2019}.
We will assume that the fitness of genotype $\mathbf{s}$ which carries allele $s_i$ on locus $i$ depends on single-locus
variations and pair-wise co-variations, that is
\begin{equation}
  \label{eq:fitness}
  F(\mathbf{s}) = F_0 + \sum_i f_is_i + \sum_{ij}f_{ij}s_{i},s_{j}
\end{equation}
The first term is an overall constant.
The second term, linear in the genome, is called \textit{additive component of fitness}.
The last term, quadratic in the genome, is called \textit{epistatic component of fitness}.
The dynamics due to Darwinian selection on the level of populations is thus
\begin{equation}
  \label{eq:fitness-dynamics}
  \frac{\partial P(\mathbf{s})}{\partial t}|_{sel}= P(\mathbf{s})\left(F(\mathbf{s})-\left<F(\mathbf{s})\right>\right)
\end{equation}
where $\left<F(\mathbf{s})\right>=\sum_{\mathbf{s}}F(\mathbf{s} P(\mathbf{s})$
is the average fitness over the population.
Evolutionary dynamics due to fitness in a fitness landscape
is thus quadratic in the distribution function
\protect\footnotemark[8]{$^8$}
\protect\footnotetext[8]{$^8~$ $P(\mathbf{s})\left<F(\mathbf{s})\right>$
is quadratic in $P(\mathbf{s})$.}.
The conditions under which the combined dynamics under mutations and
fitness satisfy detailed balance is a kind of integrability condition.
On the level of dynamics on allele frequencies this condition is known as the
existence of a Svirezhev-Shahshahani potential~\cite{Shahshahani76,Burger2000,SvirezhevPassekov2012,Huillet2017},
see also~\cite{AurellEkebergKoski2019}.

Recombination (or sex) is the mixing of genetic material between different
individuals.
In diploid organisms (such as human) every individual has two copies
of each separate component of its genetic material (chromosome),
where one comes from the father and one comes from the mother,
each of whom also has two copies, one from each grandparent.
When passing from the parents to the child
the material from the grandparents is mixed
in the process called cross-over, so that one chromosome
of the child inherited from one parent typically
consists of segments alternately taken
from the two chromosomes of that parent.

In haploid organisms the situation is both simpler since each organism
only has one copy of its genetic material, and also more complicated
since the mixing of information can happen in many different
ways.
It is convenient to postulate a dynamics like a physical collision process
\begin{eqnarray}
 \label{eq:Potts-recomb-bact}
  \frac{\partial P(\mathbf{s})}{\partial t}|_{rec}
=  r \sum_{\mathbf{\xi},\mathbf{s}'}   C(\mathbf{\xi})
  \big[&Q&{{(\mathbf{s}_1,\mathbf{s}_2) P_2(\mathbf{s}_1,\mathbf{s}_2)  }} \nonumber \\
 &-& {{Q(\mathbf{s},\mathbf{s}') P_2(\mathbf{s},\mathbf{s}') }}\big]
\end{eqnarray}
where $r$ is an overall rate of sex compared to other processes,
$Q(\mathbf{s}_1,\mathbf{s}_2))$ is the chance of individuals
$\mathbf{s}_1$ and $\mathbf{s}_2$ mating (reaction probability)
and $C(\mathbf{\xi})$ is the chance that they produce an offspring given
by a pattern $\xi$~\cite{NeherShraiman2011,Gao2019} (probability of outcome of reaction).
On the left hand side we have the single-genome
distribution function $P$, and on the right-hand side the two-genome
distribution function $P_2$; the equations are closed by a \textit{Stosszahlansatz}
\begin{eqnarray}
 \label{eq:Stosszahlansatz-1}
P_2(\mathbf{s}_1,\mathbf{s}_2) &=& P(\mathbf{s}_1)P(\mathbf{s}_2)
\end{eqnarray}
The complexities of recombination can then be accommodated by the two functions
$Q$ and $C$.
A method to infer recombination hotspots in bacterial
genomes was discussed in~\cite{Yahara-2014a},
and the issue was also discussed in~\cite{Chewapreecha-2014a},
in relation to the the same ``Maela'' data set used in~\cite{Skwark-2017a}.
Detailed descriptions of the relation between on the one hand
$(\mathbf{s},\mathbf{s}')$ and on the other
$(\mathbf{s}_1,\mathbf{s}_2)$
as parametrized by $\xi$ can be found in~\cite{NeherShraiman2011}
and \cite{Gao2019}.

\subsection{The Quasi-Linkage Equilibrium Phase}
\label{sec:QLE-phase}
The concept of Quasi-Linkage Equilibrium (QLE) and its relation to
sex was discovered by population geneticist Motoo Kimura~\cite{Kimura1956,Kimura1964,Kimura1965},
and later developed further by Richard Neher and Boris Shraiman in
two influential papers~\cite{NeherShraiman2009,NeherShraiman2011}.
We will refer to this theory of QLE as \textit{Kimura-Neher-Shraiman} (KNS) theory.
To define QLE and state the main result of KNS we must first
introduce the simpler concept
of Linkage Equilibrium (LE), which goes back to the work of Hardy and Weinberg more than a century ago~\cite{Hardy1908, Weinberg1908}.

Consider two loci $A$ and $B$ where there can be, respectively, $n_A$ and $n_B$ alleles.
The configuration of one genome
with respect to $A$ and $B$
is then $(x_A,x_B)$ where $x_A$ takes values in $\{1,\ldots,n_A\}$
and $x_B$ takes values in $\{1,\ldots,n_B\}$.
The configuration of a population of $N$ individuals
is the set $\left[(x_A^{(s)},x_B^{(s)})\right]$ where $s$ ranges from $1$ to $N$.
This set defines the empirical probability distribution with respect to $A$ and $B$ as
\begin{equation}
P_{AB}(x_A,x_B) = \frac{1}{N}\sum_{s=1}^N \mathbf{1}_{x_A^{(s)},x_A} \mathbf{1}_{x_B^{(s)},x_B},
\end{equation}
where $\mathbf{1}_{a,b}$ is the Kronecker delta.
Similarly we can define distributions over one locus as
$P_{A}(x_A) = \frac{1}{N}\sum_{s=1}^N \mathbf{1}_{x_A^{(s)},x_A}$,
and $P_{B}(x_B)$.
The distribution of genomes in a population over loci $A$ and $B$ is said to be in
\textit{Linkage Equilibrium} (LE) if the alleles $a_A$ and $x_B$ are independent under the
empirical distribution \textit{i.e.} if $P_{AB}(x_A,x_B)= P_{A}(x_A)P_{B}(x_B)$.
All other distributions are in \textit{Linkage Disequilibrium} (LD).

Specifying for completeness to the case of interest here where all loci
are biallelic and epistatic contributions to fitness is quadratic (pairwise
dependencies) \textit{Quasi-Linkage Equilibrium} is a subset of distributions in LD
where the joint distribution over loci is the Gibbs distribution in \eqref{Gibbs-distribution}.
The fundamental insight of Kimura was
that such distributions appear naturally in sexually reproducing populations
where recombination is fast~\cite{Kimura1956,Kimura1964,Kimura1965}.
In this setting epistatic contributions to fitness is a small effect since
there is a lot of mixing of genomes between individuals from one generation
to the next. The dependencies (parameters $J_{ij}$) are also small, such
that the distributions over alleles are almost independent. In other words,
the distributions in QLE which appear in KNS theory are close to being
in Linkage Equilibrium.
Nevertheless, the parameters $h_i$ and $J_{ij}$ in \eqref{Gibbs-distribution}.
are hence here \textit{consequences}
of a dynamical evolution law.

The derivation of \eqref{Gibbs-distribution}
from the dynamics described above in Section~\ref{sec:population-genetics}
has been given in the recent literature~\cite{NeherShraiman2011,Gao2019} and will be
therefore not be repeated here.
We will instead just state the most important result of KNS theory.
This is
\begin{equation} \label{KNS-eq1}
J_{ij} = \frac{f_{ij}}{rc_{ij}}
\end{equation}
where $c_{ij}$ characterizes the amount of recombination
between loci $i$ and $j$. Referring to the dynamics
\eqref{eq:Potts-recomb-bact} this quantity is defined as
\begin{equation}
  \label{eq:cij}
c_{ij} = \sum_{\mathbf{\xi}} C(\mathbf{\xi})\left(\xi_i(1-\xi_j)+(1-\xi_i)\xi_j\right)
\end{equation}
In words $c_{ij}$ is simply \textit{the probability that the alleles at
loci $i$ and $j$ were inherited from different parents.}
In most models of recombination this will depend on the
genomic distance between $i$ and $j$ such that
$c_{ij}$ will be close to zero when $i$ and $j$ are close,
and then grow to $\frac{1}{2}$ when they are far apart.

\subsection{Inferred interactions and underlying fitness}
\label{sec:interactions-fitness}
Turning around the concepts, \eqref{KNS-eq1}
can be interpreted as a \textit{inference formula of epistatic fitness from genomic data}:
\begin{equation}
\label{KNS-eq2}
f^*_{ij} = J^*_{ij} \cdot rc_{ij}
\end{equation}
where $*$ indicates inferred value.

The parameter $J^*_{ij}$ can be determined from data by DCA
while the parameters $r$ and $c_{ij}$
have to be determined by other means.
However, since the
QLE phase is characterized by
$J_{ij}$ being small, or, alternatively,
$f_{ij}$ being smaller than $rc_{ij}$,
we can make the simplifying assumption that
$c_{ij}\approx \frac{1}{2}$ for all pairs of loci
we consider.
Formula \eqref{KNS-eq2} then says that underlying
fitness parameters $f_{ij}$ are proportional to inferred Ising parameters
$J_{ij}$, where the proportionality is $r/2$.

Nevertheless, \eqref{KNS-eq2} will work also when
the variation of $c_{ij}$ is taken into account,
as long as the product $rc_{ij}$ remains smaller than
$f_{ij}$.
It is not currently clear if there exists also an extension of
\eqref{KNS-eq2} which holds also when $rc_{ij}$ is of the order
of or larger than $f_{ij}$, including for the case when
$i$ and $j$ are close (``hitch-hiking mutations'').

\subsection{Fitness inference for synthetic Ising genomic data}
\label{sec:fitness-inference-synthetic}
We here describe results obtained from simulating a finite
population using the FFPopSim software~\cite{FFPopSim}.
Partial results in the same direction were
reported in~\cite{Gao2019};
more complete results, though not using the \textit{single-time}
versions of algorithms as we will here, in~\cite{ZengAurell2020}.

\begin{figure}[ht]
\centering
\includegraphics[width=0.95\linewidth]{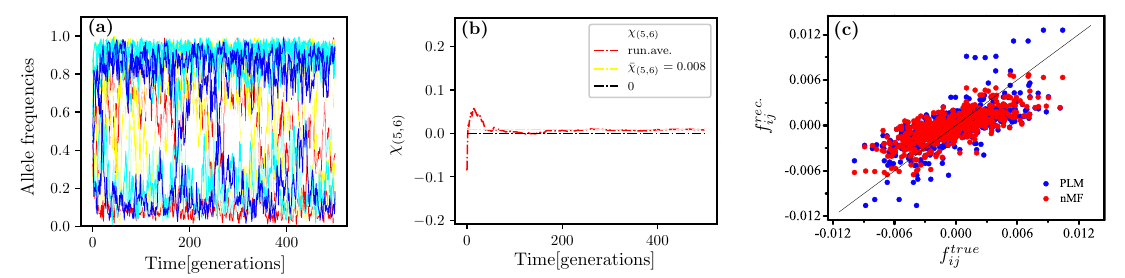}
\caption{Left panel(a): temporal behavior of all allele frequencies defined as $f_i[1]$. Data recorded every 5 generations.
Middle panel(b): an example of pairwise correlation changing with time. With finite population size, there exists strong fluctuations in the system.
Right panel(c): Scatter plot for the reconstructed against the tested fitness with DCA-nMF (red dots) and DCA-PLM (blue dots) algorithm for $J_{ij}$s.
Parameters: number of loci $L=25$, number of individuals $N=200$, mutation rate $\mu=0.01$, recombination rate $r=0.1$, crossover rate $\rho=0.5$,  standard deviation of epistatic fitness $\sigma =0.002$.}
\label{fig:time-development-scatters}
\end{figure}

In a finite population statistical genetics as described above only holds on the average; when following one population in time fluctuations of order $N^{-\frac{1}{2}}$ appear for observables such as single-locus frequencies and pair-wise loci-loci correlations.
Figure~\ref{fig:time-development-scatters}(a) and (b) reports simulations using the \texttt{FFPopSim} software for allele frequencies and a specified pair-wise loci-loci correlations that these fluctuations can in practice (in simulations) be quite large.
Figure~\ref{fig:time-development-scatters}(c) presents the reconstructed fitness by DCA-PLM (blue dots) and DCA-nMF (red dots) against the tested fitness. Both methods exhibit clear trend along the diagonal direction though with fluctuations.

The inference of fitness is governed by a set of parameters during the population evolutionary  process. The illustrated examples in simulation below contain some fixed parameters, which are the number of loci $L=25$, the number of individuals $N=500$ (to avoid the singularity of correlation matrices of single generation), the length of generations $T=500\times5$, the crossover rate $\rho=0.5$. The varied parameters are the mutation rate $\mu$, the recombination rate $r$ and the strength of fitness $\sigma$.
In the following we will discuss what one can observe by
systematically varying these three parameters.

\begin{figure}[ht]
 \centering
 \includegraphics[width=0.95\linewidth]{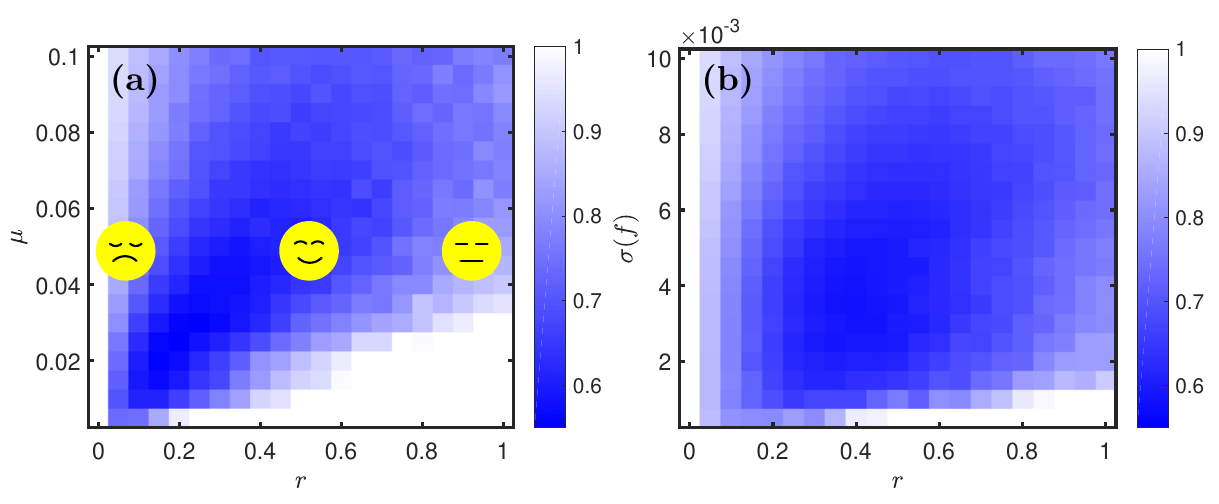}
 \caption{Phase Diagram for epistatic fitness recovery with DCA-nMF $J_{ij}$s  from the average of \textit{singletime} data.
 Left panel(a): mutation rate $\mu$ versus recombination rate $r$. For large recombination while low mutation KNS inference does not work as shown in figure \ref{fig:scatters-singletime}(c).
 However, for small $r$, the KNS inference theory does not satisfied.
 Right panel(b): epistatic fitness strength $\sigma$ with $r$.  For large recombination and very small fitness KNS inference does not work.
 }
 \label{fig:PD-singletime}
\end{figure}

Furthermore, it is of interest to see how the KNS inference theory performs by averaging the results from \textit{singletime} data. That means that we infer parameters from snapshots, and then average those inferred parameters over the time of the snapshot. In \cite{ZengAurell2020} in contrast was studied the inference using \textit{alltime} versions of the data, where inference is done only once. In figure \ref{fig:PD-singletime} we show the phase diagrams of epistatic fitness inference. The color indicates the relative root mean square error of the fitness reconstruction, where lighter color means larger error. However, the mean square error $\epsilon$ as shown in (\ref{MSE}) is used for consistence in the following scatter-plots.
Panel (a) shows the parameters mutation rate $\mu$ and  recombination rate $r$ while (b) for fitness strength $\sigma$ and $r$. Both of these have wide broad ranges of parameters where KNS theory works well for the fitness recovery. The inference phase diagrams based on $sigletime$ here are quite close to those presented in \cite{ZengAurell2020} using $alltime$.

\begin{figure}[ht]
 \centering
 \includegraphics[width=0.95\linewidth]{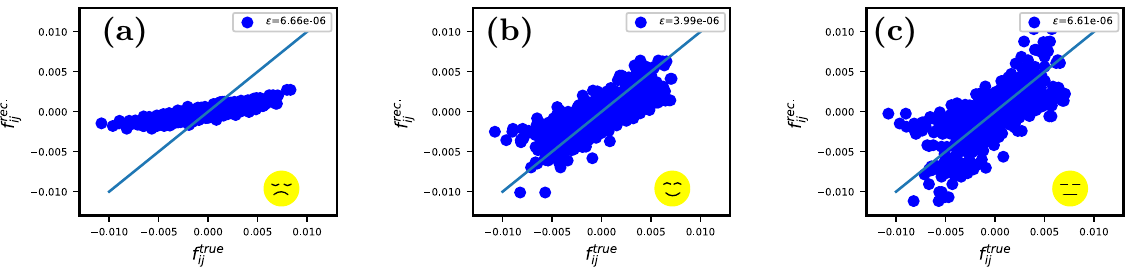}
 \caption{Scatter-plots of inferred epistatic fitness against the true fitness based the averaged results from \textit{singletime}. Here, DCA-nMF algorithm for $J_{ij}$s is utilized.
 Left panel(a) with sad face: $r=0.1$, KNS theory cannot be satisfied here.
 Right panel(b) with smiling face: $r=0.5$, inference works.
 Right panel(c) with not-that-sad face: $r=0.9$,  KNS theory works but with very heavy fluctuations.}
 \label{fig:scatters-singletime}
\end{figure}

When mutation rates are very low, the frequencies of most loci is frozen to 0 or 1 for most of the time. This is a classical fact for evolution of one single locus, as discussed above, but also holds more generally. For an evolving population simulated with the FFPopSim software it was demonstrated in~\cite{ZengAurell2020}.
In this regime fitness recovery is hence impossible as there is not enough variation.
On the other hand, the KNS inference theory does not hold for high enough $\mu$,
as one of the assumptions is that recombination is a faster process than mutations.
Thus, three points on the $\mu$-$r$ phase diagram are picked with same $\mu$ and differing $r$s, marked as sad face, smiling face and not-that-sad face respectively. The corresponding scatter plots are presented in figure \ref{fig:scatters-singletime}. As expected, KNS inference works but with very heavy fluctuations for very high $r$ but does not hold for low $r$.

\begin{figure}[ht]
 \centering
 \includegraphics[width=0.95\linewidth]{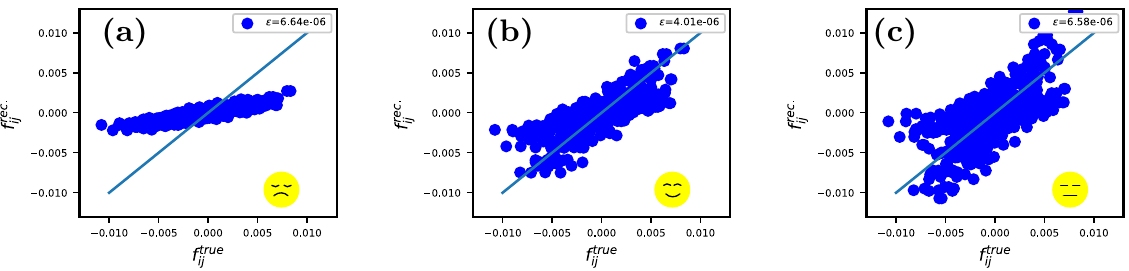}
 \caption{Corresponding scatter-plots by \textit{alltime} averages with figure \ref{fig:scatters-singletime}.
 DCA-nMF algorithm for $J_{ij}$s is used also here.
 The parameters for each sub-panel are same with those in figure \ref{fig:scatters-singletime}.
 }
 \label{fig:scatters-alltime}
\end{figure}

To see if there are differences between the inference with average over \textit{singletime} and \textit{alltime}, the corresponding scatter-plots of figure \ref{fig:scatters-singletime} (\textit{singletime}) are presented in
figure \ref{fig:scatters-alltime} (\textit{alltime}). With the parameters illustrated here, the difference between two approaches is small.

\section{Discussion and Perspectives}
\label{sec:discussion}
Inverse Ising/Potts or DCA has emerged as a powerful
paradigm of biological data analysis which has helped
to revolutionize protein structure prediction.
For the first time it has been shown to be possible
to predict protein structure from sequence, though
crucially from many similar sequences, not from a single one.
The central idea which has made this possible is to exploit
statistical dependencies encoded by a postulated
Gibbs distribution \eqref{Gibbs-distribution} of
the Ising/Potts form over sequence space.
While DCA recently has been overtaken by more complex
AI learning methods of the deep learning type,
it remains the case that it was the success of DCA
that showed this to be possible. Many other applications
have appeared, some of them in areas where AI learning methods
are not likely to succeed due to lack of training examples.

In this review we have striven to put these developments
in the context of statistical physics. On some level
a distribution over sequences must be arrived at by a
evolutionary process, which though it may be complicated,
shares aspects of non-equilibrium spin dynamics.
Indeed, these analogies have been noted for a long time,
and have been explored from both the viewpoint of
(theoretical) population genetics, and statistical physics.
We have here added the dimension of learning, how knowledge
of the type of dynamics and inference techniques
can be used together to deduce biological parameters from data.
We have also considered more direct applications
of kinetic Ising models to model the evolution of neurons
and of economic data, and how to infer connections from such data.

The main conclusions are as follows.
First, we have stressed that dynamics that does not fulfill
detailed balance can have practically arbitrarily complicated
stationary states, even if interactions is only pair-wise.
It can therefore not be the case that inverse Ising/Potts
can generally give useful information: in the wrong parameter
phase it is instead much more likely to yield garbage.
The simplest example is inference in asynchronous kinetic
Ising models discussed in Section~\ref{sec:kinetic-Ising}:
those models contain parameters (the anti-symmetric combination $J_{ij}-J_{ji}$)
that are simply not present in the Ising distribution \eqref{Gibbs-distribution}.
DCA, by whichever algorithm, therefore will never be able to find them.
Even more, the stationary distribution in such models is quite
different from \eqref{Gibbs-distribution}, and DCA is also not able
to find the symmetric combination $J_{ij}+J_{ji}$ either (unless the
anti-symmetric combination is relatively small).
On the other hand, straight-forward methods relying on inference from
time series are able to recover symmetric and anti-symmetric combinations
equally easy.
The moral of this part of our review is simple:
\textit{if you have time series data, you should use it to infer dynamic models;
it is both a more general and an easier procedure}.

Second, we have considered evolutionary dynamics in finite populations under
selection, mutations and recombination. Following the pioneering work of
Kimura and more recently Neher and Shraiman, we discussed how the high-recombination
regime leads to a distribution of the type \eqref{Gibbs-distribution},
where the parameters can be inferred by DCA.
We have noted that in the same high-recombination regime the effective
interaction parameters are small, which corresponds to the high-temperature
regime in inverse Ising. Hence inference in the high-recombination
regime is limited by finite sample noise. Given finite data inference therefore
works best in an intermediate regime, not too high recombination (because
then statistical co-variance will be too weak), and not too low recombination
(because then the Kimura-Neher-Shraiman theory does not apply).
Crucially, we have observed that though the parameters inferred by DCA
on such data are related to fitness, they are not the fitness
parameters governing the evolutionary dynamics itself. The relation is
albeit a simple proportionality, at least for pairs of loci far enough
apart on the genome, but it is not an identity.
The moral of this part of our review is thus:
\textit{if you have a theory connecting the underlying mechanism
which you want to clarify to the data which you can use,
then you are well advised to analyze the data using that theory}.

Many open questions remain in the field of DCA, out of which we will but
discuss some that are closely connected to the main thrust of our argument.
Kimura-Neher-Shraiman (KNS) theory is a huge step forward to an understanding
of what is actually inferred is such a procedure, but is obviously
only a first step.
Most directly, both Fig~\ref{fig:scatters-singletime}(a)
and Fig~\ref{fig:scatters-alltime}(a) strongly suggest
a functional relationship.
These plots were obtained in parameter regions where KNS theory cannot be expected to be valid,
and indeed it is not: the mean square error of inferred and underlying fitness
parameters is large. Since the plots suggest a functional relationship,
there should however be another theory, which at this point is unknown.
In other words, KNS is not the end, but should be the starting point
for developing theories connecting fitness (and other evolutionary parameters)
to distributions over sequences in much wider settings, and ways to learn such
parameters from sequence data.
In particular, since KNS theory is only valid when fitness
parameters $f_{ij}$ are smaller than compounded
recombination parameters $rc_{ij}$, KNS is likely not
valid for the very strongest epistatic effects which
are potentially the most interesting and biologically relevant.

Much work further deserves to be done to incorporate
further biological realism in KNS and/or its successor theories and software.
Among the many important effects (most discussed above) which
have not been taken into account in this review we list
\begin{itemize}
\item Multi-allele loci
\item Realistic mutation matrices that vary over a genome and depending on the transitions
\item Mutations that do not act on single loci \textit{i.e.} insertions and deletions (indels)
\item Other models of fitness and other distributions of \textit{e.g.} pair-wise parameters $f_{ij}$
\item More realistic models of recombination incorporating also recombination hotspots
\item More types of recombination, as appropriate for bacterial evolution
\item Effects of population growth and bottle-necks
\end{itemize}
Many kinds of simulation software has been developed in the
computational biology community, for instance
the \textit{fwdpp}~\cite{Thornton2014} software suite
used recently in~\cite{Arnold2019}.
To objective would not be to redo or replace
such software packages, but to reuse them
in the context of theory-driven inference.

One further direction important to pursue
is the effect of spatial and environmental separation, believed to be a main
mechanism behind speciation and the emergence of biological variation in general.
Its effects in models of the Wright-Fisher-Moran type were emphasized in~\cite{Blythe-2007a}.
Spatial separation would in general tend to counter-act recombination,
in that individuals which could recombine if they would meet actually are not
likely to meet.
For instance, a bacterium with one of highest known recombination rates
is the human pathogen~\textit{Helicobacter pylori} (the cause of stomach ulcers),
but two such bacteria actually can only recombine when they find themselves in the
stomach of the same host.
Strains of \textit{H. pylori} can thus be distinguished on a global scale,
and only merge when
their human host populations overlap~\cite{Thorell2017}.


\addcontentsline{toc}{chapter}{Acknowledgment}

\section*{Acknowledgment}
The material in
Sections~\ref{techniques}, \ref{sec:kinetic-Ising}
and~\ref{sec:applications-Ising} is partly based
on the PhD work of one of us
(HLZ, Aalto University, Finland, 2014).
We thank colleagues and collaborators during that time,
particularly Mikko Alava, John Hertz, Hamed Mahmoudi, Matteo Marsili and Yasser Roudi.
The material in Section~\ref{sec:population-genetics} was
motivated by many discussions in the statistical physics community
for which we thank Johannes Berg, Simona Cocco, Bert Kappen, David Lacoste, Andrey Lokhov,
R\'emi Monasson, Roberto Mulet, Andrea Pagnani, Luca Peliti,
Federico Ricci-Tersenghi,
Chris Sander,
Alex Schug,
Erik Van Nimwegen,
Martin Weigt,
Riccardo Zecchina,
and many others.
For the actual development of the material in Section~\ref{sec:population-genetics},
one of us (EA) is much indebted to a discussion with Boris Shraiman,
and collaborations and/or discussions with Magnus Ekeberg, Yueheng Lan,
Cecilia L\"ovkvist, Martin Weigt, Marcin Skwark,
Andrea Pagnani, Christoph Feinauer,
Hai-Jun Zhou, Chen-Yi Gao, Angelo Vulpiani, Fabio Cecconi,
Timo Koski, Arne Elofsson,  Daniel Falush, Kaisa Thorell, Yoshiyuke Kabashima, Jukka Corander,
Santeri Puranen, Yingying Xu, Alexander Mozeika,  R\'emi Lemoy and Onur Dikmen.
The material in
Section~\ref{sec:fitness-inference-synthetic} is the outcome of ongoing collaboration with Simona Cocco, Eugenio Mauro and R\'emi Monasson whom we thank for many fruitful discussions and suggestions on various stages of the work.
For this part we also we also thank Richard Neher for the FFPopSim software package,
and for valuable comments.
We finally thank Johannes Berg and Andrey Lokhov
for constructive comments on the manuscript.

\bibliographystyle{CPBREFERENCE}

\bibliography{main_abbr}

\end{CJK*}
\end{document}